\documentclass{article}
\usepackage{graphicx}
\usepackage{jcappub}

\title{Consistency Relation and Non-Gaussianity in a Galileon Inflation}

\author{Kosar Asadi,}

\author{Kourosh Nozari}

\affiliation{Department of Physics, Faculty of Basic Sciences,\\
University of Mazandaran,\\
P. O. Box 47416-95447, Babolsar, IRAN}

 \emailAdd{k.asadi@stu.umz.ac.ir}

\emailAdd{knozari@umz.ac.ir}

\abstract{We study a particular Galileon inflation in the light of
Planck2015 observational data in order to constraint the model
parameter space. We study the spectrum of the primordial modes of
the density perturbations by expanding the action up to the second
order in perturbations. Then we pursue by expanding the action up to the third
order and find the three point correlation functions to find the
amplitude of the non-Gaussianity of the primordial perturbations in this setup.
We study the amplitude of the non-Gaussianity both in equilateral
and orthogonal configurations and test the model
with recent observational data. Our analysis shows that for some
ranges of the non-minimal coupling parameter, the model is
consistent with observation and it is also possible to have large
non-Gaussianity which would be observable by future improvements
in experiments. Moreover, we obtain the tilt of the tensor power
spectrum and test the standard inflationary consistency relation
($r=-8n_T$) against the latest bounds from the Planck2015
dataset. We find a slight deviation from the standard consistency
relation in this setup. Nevertheless, such a deviation seems not
to be sufficiently remarkable to be detected confidently.}


\begin{document}

\maketitle

%%%%%%%%%%%%%%%%%%%%%%%%%%%%%%%%%%%%%%%%%%%%%%%%%%%%%%%%%%%%%%%%%%%%%%%%%%%%%%%%%%%%%%%%%%%%%%%%%%%%%%%%%%%%%%
%%%%%%%%%%%%%%%%%%%%%%%%%%%%%%%%%%%%%%%%%%%%%%%%%%%%%%%%%%%%%%%%%%%%%%%%%%%%%%%%%%%%%%%%%%%%%%%%%%%%%%%%%%%%%%

\section{Introduction}

Theory of cosmological inflation is basically related to a
quasi-de Sitter universe, a homogeneous and isotropic universe
that expands almost exponentially fast, with a nearly constant
event horizon. This paradigm is known as the most successful
theory to date that describes primordial density perturbations of
the universe. These primordial perturbations are arisen from the
quantum behavior of both the spacetime metric and the scalar field
which derives inflation  \cite{1,2,3,4,5,6,7,8}. Simple
inflationary models predict (for instance, via study of the Cosmic
Microwave Background (CMB) radiation) nearly scale invariant,
adiabatic and Gaussian perturbations \cite{9,10,11,12,13}.
However, some degree of non-linearities must be present at some
level in all inflation models, since non-linearity is the inherent
property of gravity in essence and also most of the inflationary
scenarios have interacting or self-interacting potentials
\cite{14,15,16}. Non-Gaussianity in the perturbation's mode could
be relatively large in some situations (see for instance
\cite{17,18,19,20,21,22,23,24,25,26}). It is now well-known that
large non-Gaussianities naturally follow inflationary scenarios
with higher derivative terms \cite{15,27}. One might well hope to
detect this large non-Gaussianities by future improvements in
experiments, which would provide another key observable to
constraint or confirm particular inflation models and also the
underlying high energy theories from which they are originated
\cite{15}.

While higher derivative theories predict some levels of
non-Gaussianities, these theories are typically plagued by ghost
instabilities \cite{28,29}. In Horndeski construction these
instabilities are avoided \cite{30}. The most general Horndeski
single-field Lagrangian leading to second-order equations of
motion covers a wide variety of gravitational theories with one
scalar degree of freedom. Among these theories one could mention
standard slow-roll inflationary model \cite{2,3,31,32,33,34},
non-minimally coupled scenarios \cite{35,36,37}, Brans-Dicke
theories \cite{38} (which include $f(R)$ gravity \cite{39}),
Galileon inflation \cite{27,40,41,42}, k-inflation \cite{43,44}
and field derivative coupled to gravity \cite{45,46,47}). As a
comprehensive study of the most general non-canonical and
non-minimally coupled single field inflation models which yield
second-order field equations, one can be referred to Kobayashi
\emph{et al.} \cite{48}. In \cite{48} the authors have presented
the most general extension of the Galileons which are no longer
based on a symmetry argument (since such a symmetry must be
practically broken to terminate inflation and reheat the
Universe). In order to give the stability analysis and the power
spectrum of the primordial fluctuations in the generalized
G-inflation, they have illustrated the generic behavior of the
inflationary background and studied the nature of primordial
perturbations at \emph{linear} order, by obtaining the most
general second-order action for both scalar and tensor density
perturbations. On the other hand, these complex scenarios are
capable to provide a dark energy component for late time dynamics
of the universe \cite{49,50,51,52,53,54,55,56}. In these extended
scenarios, the fields are considered to be non-minimally coupled
to the background curvature leading to interesting cosmological
outcomes in both dark energy and inflation eras. Since operators
coming out from the non-minimal coupling are non-renormalizable,
the unitarity bound of the theory during inflation era is violated
\cite{57,58,59,60}. One desires to avoid this unitarity violation
to find a framework that the Higgs boson would behave like a
primordial Inflaton. In this regard, considering non-minimal
coupling between the derivatives of the scalar fields and
curvature can be a solution \cite{61,62,63,64,65,66}. This
scenario can be regarded as a subset of the most general
scalar-tensor theories. In this respect, studying the second-order
gravitational theories in connection with Galileon gravity
\cite{67,68,69} has attracted much attention recently. These
models can be considered generally as rediscovery of Horndeski
construction and its detailed features. To see how studies have
been improved in this direction we refer to Refs.
\cite{39,41,70,71,72,73,74,75,76,77,78,79,80,81,82,83,84,85,86,87,88,89,90}.
Galileon models with non-minimal derivative coupling have also
frequently received attention as models of dark energy, because
they can describe an effective short-distance theory associated
with a modification of gravity on large scales
\cite{70,71,91,92,93}.

In this paper, we study a particular Galileon inflation which is a
radiatively stable higher derivative model of inflation. This
model is determined by a finite number of relevant operators which
are protected by a covariant generalization of the Galileon shift
symmetry (see for instance \cite{27,94}). By comparing the
obtained values of the observable parameters (such as the scalar
spectral index and tensor-to-scalar ratio) in this setup with
observation, viability of the model at hand can be tested to see
whether it is successful or not. The combined
WMAP9+eCMB+BAO+H$_{0}$ data results in the constraints $r < 0.13$
and $n_s = 0.9636 \pm 0.0084$ \cite{95}, while the joint
Planck2013+WMAP9+BAO data expresses the conditions  $r < 0.12$ and
$n_s = 0.9643 \pm 0.0059$ \cite{96}. The more recent study of the
Planck team has constrained the scalar spectral index and
tensor-to-scalar ratio as $r < 0.099$ and $n_s=0.9652 \pm 0.0047$,
from Planck TT, TE, EE+low P+WP data \cite{97,98,99}. We note that
Planck TT, TE, EE+lowP refers to the combination of the likelihood
at $l>30$ using TT, TE, and EE spectra and also the low $l$
multipole temperature polarization likelihood. Therefore, in order
to compare our model with observational data, we study the
behavior of the tensor-to-scalar ratio, $r$, versus the scalar
spectral index, $n_s$, in the background of the Planck TT, TE,
EE+low P data and find some constraints on the parameters space of
the model. We focus our attention on several choices of functions
of the scalar field in the action including $K(\phi)$,
$\gamma(\phi)$ and the potential, $V(\phi)$. Furthermore, in each
case, we analyze the deviation from the standard slow-roll
consistency relation $r=-8n_T$ due to the effect of additional
terms in the action. We note that the issue of consistency
relation and its status in Galileon and DBI Galileon inflation are
studied in \cite{41,42,48,100,101}. In Ref. \cite{100}, the
authors by considering restricted class of Galilean inflation
models with Lagrangian ${\cal{L}}(\phi,X,\Box{\phi})=\frac{X^n
\Box{\phi}}{M^{4n-1}}-V(\phi)$ have shown that depending on the
value of the parameter $n$, one can have both $r>-8n_{T}$ or
$r<-8n_T$, in spite of the fact that the speed of sound is
subluminal in both the cases. As we will show a slight deviation
from the standard consistency relation can be obtained in our
setup too.

We study also non-Gaussian feature of the perturbations numerically by focusing on the
behavior of the orthogonal versus equilateral configuration in the background of the observational data.
Non-Gaussianity will be detectable essentially by observation if
the bispectrum, which is transformation of the three-point
correlator in Fourier space, is of the order of the square of the
power spectrum. Using different types of functions $K(\phi)$,
$\gamma(\phi)$ and the potential $V(\phi)$, we show that for some
ranges of the non-minimal coupling parameter, our model is
consistent with observation and it is also possible to have large
non-Gaussianity in this setup. In other words, we show that the non-Gaussianity
of the primordial density perturbation generated during an epoch
of Galileon inflation is a particularly powerful observational
probe of these models. We also note that large non-Gaussianities
would be observable by future improvements in experiments and in
this respect this would be an important result in our
study.

The paper is organized as follows: After describing the action for
the Galileon inflationary model and deriving the main equations of
the model in section II, we investigate the linear perturbation of
the model in Section III. We expand the action up to the second
order in perturbations and derive the two-point correlation function
which gives the amplitude of the scalar perturbation and its
spectral index. Furthermore, we study the tensor part of the
perturbed metric and obtain the tensor perturbation and its spectral
index. We also show that in the model under consideration, the
consistency relation of the standard inflation is modified. In Sec.
IV, we expand the action up to the cubic order in perturbation to
investigate nonlinear perturbation in the model. In this regard, the
amplitude of the non-Gaussianity is derived in the equilateral and
orthogonal configurations and with the equilateral limit,
$k_1=k_2=k_3$. In this way, we focus on the possibility of deriving
a large amount of non-Gaussianity in this particular Galileon inflation and
obtain some conditions to have large non-Guassianity. In Sec. V, we
test our Galileon inflationary model in confrontation with the
recently released observational data and find constraints on the
parameters space of the model. We also test deviation from the
standard consistency relation in each case. Finally, we give our
summary and conclusions in Sec. VI.

\section{The Model}

The unique 4-dimensional action of the most general class of
scalar-tensor theories which contains both first and second
derivatives of the scalar field, and leads to the second order
equations of motion for both the metric and scalar field, is given
as follows
\begin{eqnarray}\label{action}
S=\int d^{4}x\sqrt{-g}\left(\frac{M_{pl}^2}{2}R+{\cal{L}}\right)\,,
\end{eqnarray}
where $R$ is the Ricci scalar and represents the Einstein-Hilbert
action of General Relativity and
\begin{eqnarray}\label{lag}
{\cal{L}}\equiv{\cal{L}}(\phi,\nabla\phi,\nabla\nabla\phi)=\sum_{i=2}^{5}{\cal{L}}_{i}\,,
\end{eqnarray}
which collects the higher order corrections to General Relativity.
We should note that, in order to constitute a well-defined,
predictive description of an inflationary phase, there are some
conditions that the Lagrangian of this generated model,
${\cal{L}}$, must satisfy. An arbitrary action of the form
(\ref{action}) contains second and higher-order derivatives of the
inflaton field, and therefore results in equations of motion of
third-order or higher. Such theories do not usually possess a
well-defined Cauchy problem \cite{27,102}, (unless presenting an
infinite number of derivatives), and this leads to appearance of
ghost states which causes a pernicious loss of unitarity at the
quantum level. For this reason, we should restrict our attention
to special choices of ${\cal{L}}$ which lead to second-order
equations of motion \cite{103,104,105}. In Ref. \cite{106} it has
been shown that choosing ${\cal{L}}=G(X,\phi) \Box{\phi}$ leads to
second-order equations of motion for any choice of $G(X,\phi)$,
and also unitary evolution as a quantum field theory. Thus, we
consider such models as candidates for an inflationary action of
the form (\ref{action}) and define the following expressions for
${\cal{L}}_i$
\begin{eqnarray}\label{3}
{\cal{L}}_{2}=K(\phi,X)\,,
\end{eqnarray}
\begin{eqnarray}\label{4}
{\cal{L}}_{3}=-G_{3}(\phi,X)\Box\phi\,,
\end{eqnarray}
\begin{eqnarray}\label{5}
{\cal{L}}_{4}=G_{4}(\phi,X)R+G_{4,X}\Big[(\Box\phi)^2-(\nabla_{\mu}\nabla_{\nu}\phi)^2\Big]\,,
\end{eqnarray}
\begin{eqnarray}\label{6}
{\cal{L}}_{5}=G_{5}(\phi,X)G_{\mu\nu}(\nabla^{\mu}\nabla^{\nu}\phi)
- \frac{1}{6}G_{5,X}\Big[(\Box\phi)^3-
3(\Box\phi)(\nabla_{\mu}\nabla_{\nu}\phi)(\nabla^{\mu}\nabla^{\nu}\phi)+\hspace{2.5cm}\\\nonumber
2(\nabla^{\mu}\nabla_{\alpha}\phi)
(\nabla^{\alpha}\nabla_{\beta}\phi)(\nabla^{\beta}\nabla_{\mu}\phi)\Big]\,.
\end{eqnarray}
Here, $\phi$ is a scalar field coupled with gravity, $K(\phi,X)$ and
$G_{n}(\phi,X)$ ($n=3,4,5$) are ordinary functions in terms of the
scalar field and $
X=-\frac{1}{2}g^{\mu\nu}\nabla_{\mu}\phi\nabla_{\nu}\phi\,$,
$R_{\mu\nu}$ is the Ricci tensor and $G_{\mu\nu}$ is the Einstein
tensor ($G_{\mu\nu}=R_{\mu\nu}-\frac{1}{2}g_{\mu\nu}R$). ${G_{,X}}$
denotes derivative of $G$ with respect to $X$. In this paper, we
study how inflation can be realized in theories of the form
(\ref{action}).

Setting ${\cal{L}}_2\neq0$ and $G_{n}(\phi,X)=0$ in the general
action results the k-inflation model and in order to obtain the
Lagrangian of G-inflation, we must set ${\cal{L}}_2,G_3\neq0$ and
$G_4=G_5=0$. However, by taking into account nonzero Lagrangians
${\cal{L}}_4$ and ${\cal{L}}_5$, one can cover a wide variety of
gravitational theories such as scalar-tensor theories, field
derivative couplings with gravity, Galileon gravity and
higher-curvature gravity including Gauss-Bonnet and f(R) theories.

In what followes we consider the case
\begin{eqnarray}\label{7}
K(X,\phi)={\cal{K}}(\phi)X-V(\phi)\,,
\end{eqnarray}
\begin{eqnarray}\label{8}
G_{3}(\phi,X)=-\gamma(\phi)X\,,
\end{eqnarray}
\begin{eqnarray}\label{9}
G_{4}(\phi,X)=\frac{1}{2}\Big(\xi\phi^2+\frac{X^2}{\mu^2}\Big)\,,
\end{eqnarray}
and
\begin{eqnarray}\label{10}
G_{5}(\phi,X)=0\,,
\end{eqnarray}
where ${\cal{K}}$ and $\gamma$ are ordinary functions of the scalar
field, $V$ is the potential, $\mu$ is a constant and the first term
in $G_4$ shows an explicit non-minimal coupling of the scalar field
with the Ricci scalar with $\xi$ being the non-minimal coupling
parameter. Let us assume a spatially flat Friedmann-Robertson-Walker
spacetime
\begin{eqnarray}\label{11}
ds^{2} = -dt^{2} + a^{2}(t)\delta_{ij}dx^{i}dx^{j}\,.
\end{eqnarray}
Variation of the action with respect to the scalar field gives the following
equation of motion
\begin{eqnarray}\label{12}
\left(-4\gamma_{,\phi}X-{\cal{K}}+6\gamma H\dot{\phi}-\frac{27
H^2\dot{\phi}^2}{\mu^2}\right)\ddot{\phi}+
\left(3\gamma\dot{\phi}^2+6\xi\phi-\frac{18
H\dot{\phi}^3}{\mu^2}\right)\dot{H}-\hspace{1.5cm}\\\nonumber3\left(\frac{12
H^2 X}{\mu^2}+{\cal{K}}\right)H\dot{\phi}+
\left(9H^2\gamma-\frac{1}{2}{\cal{K}}_{,\phi}-
\gamma_{,\phi\phi}X\right)\dot{\phi}^2+12H^2\xi\phi-V_{,\phi}=0\,,
\end{eqnarray}
and varying the action with respect to the metric leads to the
following Friedmann equations
\begin{eqnarray}\label{13}
3H^2\left(M_{pl}^2+\xi\phi^2+\frac{ X^2}{\mu^2}\right)-
\left(\gamma_{,\phi}X+\frac{12
H^2\dot{\phi}^2}{\mu^2}+\frac{1}{2}{\cal{K}}\right)\dot{\phi}^2+6\xi
H\phi\dot{\phi}-V=0\,,
\end{eqnarray}
and
\begin{eqnarray}\label{14}
-2\dot{H}\left(\frac{3 X^2}{\mu^2}-\xi\phi^2-M_{pl}^2\right)=2\xi
H\phi\dot{\phi}- \left(2\gamma_{,\phi}X+\frac{9
H^2\dot{\phi}^2}{\mu^2}+{\cal{K}}+2\xi\right)\dot{\phi}^2-\\\nonumber
\left(2\xi\phi-\frac{6
H\dot{\phi}^3}{\mu^2}+\gamma\dot{\phi}^2\right)\ddot{\phi}\,,
\end{eqnarray}
where a dot represents the derivative with respect to the cosmic
time $t$. In this
inflationary model, the slow-roll parameters which are defined as
$\epsilon\equiv-\frac{\dot{H}}{H^2}$ and
$\eta\equiv-\frac{1}{H}\frac{\ddot{H}}{\dot{H}}$ are given by the
following expression
\begin{eqnarray}\label{15}
\epsilon=\frac{{\cal{F}}}{H^2\Big(M_{pl}^2+\xi\phi^2-3\frac{
X^2}{\mu^2}\Big)}\,,
\end{eqnarray}
\begin{eqnarray}\label{16}
\eta=2\epsilon-\frac{\dot{\cal{F}}}{H^3
\epsilon\Big(M_{pl}^2+\xi\phi^2-3\frac{ X^2}{\mu^2}\Big)}+
\frac{2{\cal{F}}\dot{H}}{H^4\epsilon\Big(M_{pl}^2+\xi\phi^2-3\frac{
X^2}{\mu^2}\Big)}-\frac{{\cal{F}}\Big(6\frac{X}{\mu^2}
\dot{\phi}\ddot{\phi}-2\xi\phi\dot{\phi}\Big)}{H^3\epsilon\Big(M_{pl}^2+\xi\phi^2-3\frac{
X^2}{\mu^2}\Big)^2}\,,\hspace{0.8cm}
\end{eqnarray}
where parameter ${\cal{F}}$ is defined as
\begin{eqnarray}\label{17}
{\cal{F}}={\cal{K}}X-\frac{12 X^2}{\mu^2\dot{\phi}}\ddot{\phi}
-3H\gamma X\dot{\phi}+2\gamma_{,\phi}X^2+\frac{18 H^2X^2}{\mu^2}-\xi
H\phi\dot{\phi}+ 2\xi X+\gamma
X\ddot{\phi}+\xi\phi\ddot{\phi}\,.\hspace{1cm}
\end{eqnarray}

Since during the inflationary era, the evolution of the Hubble
parameter is so slow, the conditions $\epsilon\ll1$ and $\eta\ll1$
are satisfied in this regime and whenever one of these two
conditions for the slow-varying parameters violates, the inflation
phase terminates.

The minimal number of e-folds
during inflationary regime is given by
\begin{eqnarray}\label{18}
N=\int_{t_{*}}^{t_{f}}{H dt}
\end{eqnarray}
and within the slow-roll limit, ($\ddot{\phi}\ll\,
 |{3H\dot{\phi}}|$ and $\dot{\phi}^2\ll V(\phi)$), it takes the
following form in our setup
\begin{eqnarray}\label{19}
N=\int_{\phi_{*}}^{\phi_{f}}{\frac{{\cal{K}}V+2\xi\phi
V_{,\phi}-2\xi^2 R\phi^2}{(M_{pl}^2+\xi\phi^2)\left(\xi
R\phi-V_{,\phi}\right)} d\phi}
\end{eqnarray}
Here, $\phi_{*}$ denotes the value of the inflaton field at the
horizon crossing of the universe scale and $\phi_{f}$ determines
it's value at the time of exit from inflationary phase. Up to this
point, we obtained the main equations of this extended inflationary
model. In order to test this model, we proceed by studying the
linear perturbation of the primordial fluctuations. To this end, we
study the spectrum of perturbations which are produced by quantum
fluctuations of the fields about their homogeneous background
values.

\section{Linear Perturbation}
Now, we study linear perturbations of the model introduced in the
previous section. These perturbations arise from the quantum
behavior of both the space-time metric and the scalar field around
the homogeneous background solutions.

At first, we should expand the action of the model up to the
second order of small fluctuations. In this regard, it is
convenient to use the Arnowitt-Deser-Misner (ADM) formalism in
which, by choosing a suitable gauge, one can eliminate one extra
degree of freedom of perturbations from the beginning of the
calculation \cite{107}. The space-time metric in this formalism is
given by
\begin{equation}\label{20}
ds^{2}=-N^{2}dt^{2}+h_{ij}\big(dx^{i}+N^{i}dt\big)\big(dx^{j}+N^{j}dt\big)\,,
\end{equation}
where $N$ and $N^i$ are the lapse function and the shift vector,
respectively. One can obtain the general perturbed form of the
metric by expanding the lapse function as $N = 1+{\cal{B}}$ and
the shift vector as $N^i = \delta^{ij}\partial_{j}\zeta+v^{i}$ in
which ${\cal{B}}$ and $\zeta$ are 3-scalar and $v^{i}$ is a vector
satisfying the condition $v^i_{,i} = 0$ \cite{108,109}. We note
that, since the second order perturbation is multiplied by a
factor which is vanishing using the first order solution, it is
sufficient to compute $N$ or $N^i$ up to the first order. The
contribution of the third order term also vanishes because of
multiplying by a constraint equation at the zeroth order obeying
the equations of motion \cite{13,29,110}. The coefficient $h_{ij}$
should be written as
$h_{ij}=a^2[(1-2\varTheta)\delta_{ij}+2{\cal{T}}_{ij}]$, with
$\varTheta$ being the spatial curvature perturbation and
${\cal{T}}_{ij}$ being a spatial shear 3-tensor which is both
symmetric and traceless. Finally, the above perturbed metric
(\ref{20}) takes the form
\begin{eqnarray}\label{21}
ds^2=-(1+2{\cal{B}})dt^2+2a(t)\zeta_{,i}dtdx^i+
a^2(t)\Big[(1-2\varTheta)\delta_{ij}+2{\cal{T}}_{ij}\Big]dx^{i}dx^{j}\,.
\end{eqnarray}
Now, in order to study the scalar perturbation of the theory, one
can choose the uniform-field gauge in which $\delta\phi=0$, which
fixes the time-component of the gauge-transformation vector
$\xi^{\mu}$, and also the gauge ${\cal{T}}_{ij}=0$. Considering
the scalar part of the perturbations at the linear level and
within the uniform-field gauge, finally the perturbed metric can
be rewritten as \cite{108,109,111}
\begin{eqnarray}\label{22}
ds^{2}=-\big(1+2{\cal{B}}\big)dt^{2}+2a(t)\zeta_{,i}dx^{i}dt +
a^{2}(t)\big(1-2\varTheta\big)\delta_{ij}dx^{i}dx^{j}\,.
\end{eqnarray}
Upon integrating out the auxiliary fields ${\cal{B}}$ and
$\zeta$ (see Appendix \textbf{A}), one obtains for the quadratic action
\begin{eqnarray}\label{26}
\emph{S}_2=\int dt d^3x a^3{\cal{X}}
 \left[\dot{\varTheta}^2-\frac{c_s^2}{a^2}(\partial{\varTheta})^2\right]\,,
\end{eqnarray}
where the parameters ${\cal{X}}$ and $c_s^2$ (known as the sound
speed squared), are expressed as

\begin{eqnarray}\label{27}
{\cal{X}}=\frac{4(M_{pl}^2+2G_4-4XG_{4,X})^2q_{2}}{3q_{1}^2}+3(M_{pl}^2+2G_4-4XG_{4,X})\,,
\end{eqnarray}
and
\begin{eqnarray}\label{28}
c_{s}^2=3\Big[2H(M_{pl}^2+2G_4-4XG_{4,X})^2q_{1}-(M_{pl}^2+2G_4)q_{1}^2+\hspace{2cm}\\\nonumber
4(M_{pl}^2+2G_4-4XG_{4,X})\frac{d}{dt}(M_{pl}^2+2G_{4}-4XG_{4,X})q_1-
2(M_{pl}^2+2G_4-4XG_{4,X})^2\frac{d}{dt}q_1\Big]\times\\\nonumber
\left[(M_{pl}^2+2G_4-4XG_{4,X})(4(M_{pl}^2+2G_4-4XG_{4,X})q_2+9q_1^2)\right]^{-1}\,,\hspace{1cm}
\end{eqnarray}
where $q_1$ and $q_2$ are given by
\begin{eqnarray}
q_{1}=2H(M_{pl}^2+2G_4)-2X\dot{\phi}G_{3,X}-16H(XG_{4,X}+X^2G_{4,XX})+2\dot{\phi}(G_{4,\phi}+2XG_{4,\phi
X})\,,
\end{eqnarray}
and
\begin{eqnarray}
q_{2}=-9H^2(M_{pl}^2+2G_4)+3(X K_{,X}+2X^2 K_{,XX})+18H
\dot{\phi}(2XG_{3,X}+X^2G_{3,XX})-\\\nonumber
6X(G_{3,\phi}+XG_{3,\phi
X})+18H^2(7XG_{4,X}+16X^2G_{4,XX}+4X^3G_{4,XXX})-\hspace{0.5cm}\\\nonumber
18H\dot{\phi}(G_{4,\phi}+5XG_{4,\phi X}+2X^2G_{4,\phi
XX})\,.\hspace{2cm}
\end{eqnarray}
More details in deriving these type of equations can be found in
Refs. \cite{16,20,29,88}.

Let us now proceed with the spatial curvature perturbation,
$\varTheta$. Varying the action (\ref{26}) yields the following
equation of motion
\begin{eqnarray}\label{29}
\ddot{\varTheta}+\left(3H+\frac{\dot{\cal{X}}}{{\cal{X}}}\right)\dot{\varTheta}+c_s^2\frac{k^2}{a^2}\varTheta=0\,.
\end{eqnarray}
Up to the lowest order of the slow-roll approximation, the solution
of this equation is given by the following expression
\begin{eqnarray}\label{30}
\varTheta=\frac{iHe^{-ic_s^2k\tau}}{2(c_sk)^{3/2}\sqrt{\cal{X}}}\left(1+ic_sk\tau\right)\,.
\end{eqnarray}
In order to study the power spectrum of the curvature perturbation
for the model at hand, one needs to obtain the two-point correlation
function. The two-point correlation function of curvature
perturbations can be derived by obtaining the vacuum expectation
value of $\varTheta$ at $\tau=0$ (which corresponds to the end of
the inflation phase), as
\begin{equation}\label{31}
\langle 0|\varTheta(0,\textbf{k}_{1})\, \varTheta
(0,\textbf{k}_{2})|0\rangle=\frac{2\pi^{2}}{k^3}\big(2\pi\big)^{3}
{\cal{A}}_{s}\delta^{3}\big(\textbf{k}_{1}+\textbf{k}_{2}\big)\,.
\end{equation}
Here, ${\cal{A}}_s$ is called the power spectrum of the scalar
perturbations which is defined as
\begin{equation}\label{32}
{\cal{A}}_{s}=\frac{H^2}{8\pi^2{\cal{X}}c_s^3}\,.
\end{equation}
The scalar spectral index of the perturbations at $c_s k=aH$ (which
corresponds to the time of Hubble crossing, with $k$ being the wave
number) is defined as follows
\begin{eqnarray}\label{33}
n_s-1=\frac{d \ln{{\cal{A}}_s}}{d\ln{k}}\,.
\end{eqnarray}
Before obtaining the scalar spectral index in our model, it is
convenient to introduce the following parameter
\begin{eqnarray}\label{34}
{\cal{E}}_s=\frac{c_s^2{\cal{X}}}{M_{pl}^2+
\xi\phi^2+\frac{X^2}{\mu^2}}\,,
\end{eqnarray}
which using (\ref{27}) and (\ref{28}), we can rewrite it as
\begin{eqnarray}\label{35}
{\cal{E}}_s=\epsilon+\frac{(\xi\phi-\gamma
X)\dot{\phi}+8\frac{X^2}{\mu^2}}{H\Big(M_{pl}^2+\xi\phi^2+\frac{X^2}{\mu^2}\Big)}+{\cal{O}}(\epsilon^2)\,.
\end{eqnarray}
Then equation (\ref{32}) can be rewritten as follows
\begin{eqnarray}\label{36}
{\cal{A}}_{s}=\frac{H^2}{8\pi^2\Big(M_{pl}^2+\xi\phi^2+
\frac{X^2}{\mu^2}\Big) {\cal{E}}_s c_s}\,.
\end{eqnarray}
Finally the scalar spectral index in our model can be derived as
follows
\begin{eqnarray}\label{37}
n_s-1=-2\epsilon-\frac{1}{H}\frac{\dot{c_s}}{c_s}-
\frac{1}{H}\frac{\dot{{\cal{E}}_s}}{{\cal{E}}_s}
-\frac{1}{H}\frac{d}{dt}\ln{\Big(M_{pl}^2+\xi\phi^2+
\frac{X^2}{\mu^2}\Big)}\,.
\end{eqnarray}
Any deviation of $n_s$ from the unity confirms the scale dependance
of the perturbations.

Let us now proceed further by studying the amplitude of the tensor
perturbation and its spectral index. In order to obtain the power
spectrum of the gravitational waves in this setup, we study the
tensor perturbations of the form
\begin{eqnarray}\label{38}
ds^2=-dt^2+a^2(t)(\delta_{ij}+h^{TT}_{ij})dx^{i}dx^{j}\,,
\end{eqnarray}
which is the tensor part of the perturbed metric (\ref{21}).
$h^{TT}_{ij}$ can be decomposed in terms of the two polarization
modes as follows
\begin{eqnarray}\label{39}
h^{TT}_{ij}=h_{+}e^{+}_{ij}+h_{\times}e^{\times}_{ij}
\end{eqnarray}
where $e^{(\times,+)}_{ij}$ are symmetric, traceless and
transverse tensors. In this case, the
second-order action for the tensor mode can be written as follows
\begin{eqnarray}\label{44}
\emph{S}_T=\int
dtd^{3}xa^3{\cal{X}}_{T}\Bigg[\dot{h}^2_{(+)}-\frac{c_T^2}{a^2}(\partial{h_{(+)}})^2
+\dot{h}^2_{(\times)}-\frac{c_T^2}{a^2}(\partial{h_{(\times)}})^2\Bigg]\,,
\end{eqnarray}
where the parameters ${\cal{X}}_{T}$ and $c_T^2$ are defined as
\begin{eqnarray}\label{45}
{\cal{X}}_{T}=\frac{1}{4}\left(M_{pl}^2+\xi\phi^2-3\frac{X^2}{\mu^2}\right)\,,
\end{eqnarray}
and
\begin{eqnarray}\label{46}
c^2_{T}=\frac{M_{pl}^2+\xi\phi^2+\frac{X^2}{\mu^2}}{M_{pl}^2
+\xi\phi^2-3\frac{X^2}{\mu^2}}\,,
\end{eqnarray}
respectively. Following the strategy performed for the scalar
perturbations, the amplitude of the tensor perturbations is obtained
as
\begin{eqnarray}\label{47}
{\cal{A}}_T=\frac{H^2}{2\pi^2{\cal{X}}_Tc^3_{T}}\,.
\end{eqnarray}
Using the definition of the spectral index of the gravitational
waves
\begin{eqnarray}\label{48}
n_T=\frac{d\ln{\cal{A}}_T}{d \ln{k}}\,,
\end{eqnarray}
we obtain the following expression for $n_T$
\begin{eqnarray}\label{49}
n_T=-2\epsilon-\frac{2}{\left(M_{pl}^2+\xi\phi^2+\frac{X^2}
{\mu^2}\right)}\left[\frac{\xi\phi\dot{\phi}}{H}+\frac{2 X^2
\ddot{\phi}}{\mu^2H\dot{\phi}}\right],\hspace{0.6cm}
\end{eqnarray}
which using (\ref{35}) can be rewritten as
\begin{eqnarray}\label{50}
n_T=-2{\cal{E}}_s-\frac{2\gamma X\dot{\phi}-16 H
\frac{X^2}{\mu^2}}{H\left(M_{pl}^2+\xi\phi^2+
\frac{X^2}{\mu^2}\right)}+{\cal{O}}(\epsilon^2)\,,
\end{eqnarray}
equivalently. The tensor-to-scalar ratio is another important
inflationary parameter which provides some information about the
perturbation and is defined as
\begin{eqnarray}\label{51}
r=\frac{{\cal{A}}_T}{{\cal{A}}_s}\,.
\end{eqnarray}
This parameter takes the following form in our setup,
\begin{eqnarray}\label{52}
r=16\frac{c_s}{c_T}{\cal{E}}_s\,.
\end{eqnarray}
Using equation (\ref{50}) in (\ref{52}), leads to the following
expression for $r$
\begin{eqnarray}\label{53}
r\simeq-8c_s\left[n_T+\frac{2\gamma X\dot{\phi}-16 H
\frac{X^2}{\mu^2}}{H\left(M_{pl}^2+\xi\phi^2+\frac{X^2}{\mu^2}\right)}\right]\,.
\end{eqnarray}
As we know, there is a relation between $r$ and $n_s$ in the
standard inflationary model as $r=-8 c_{s} n_{T}$ which is known as
the consistency relation. Equation (\ref{53}) shows that in our
model the consistency relation of the standard inflation is
modified, which is because of the presence of the terms $G_3$ and
$G_4$ in our action. Up to now, we obtained the primordial
fluctuations in linear order. In the next section, we explore the
non-Gaussianity of the density perturbations using the nonlinear
perturbations.

\section{Nonlinear Perturbations and Non-Gaussianity}

In this section, we study the non-Gaussianity of the primordial
density perturbation which is another important aspect of an
inflationary model. In order to compute the amount of
non-Gaussianity in a specific inflation model, we should go beyond
the linear order perturbation theory. In other words, the two-point
correlation function of the scalar perturbations gives no
information about the non-Gaussian distribution of the primordial
perturbations in the model. Thus, one has to study higher order
correlation functions. Since for a Gaussian perturbation, all odd
$n$-point correlators vanish and the higher even $n$-point
correlation functions are expressed in terms of sum of products of
the two-point functions, so, we should study the three-point
correlation function in order to explore the non-Gaussianity of the
density perturbations. To this end, we should expand the action of
the model up to the cubic order in the small fluctuations around the
homogeneous background solution.

We should remind that cubic terms obtained in this manner lead to a
change both in the ground state of the quantum field and
nonlinearities in the evolution. After expanding the action (1) up
to the third order in perturbation, one should eliminate the
perturbation parameter ${\cal{B}}$ in the expanded action. By
introducing an auxiliary field ${\cal{Q}}$ which satisfies the
following expressions
\begin{eqnarray}\label{54}
\zeta=\left[\frac{a^2}{M_{pl}^2+\xi\phi^2-3\frac{X^2}{\mu^2}}\right]{\cal{Q}}+
\left[\frac{M_{pl}^2+\xi\phi^2-3\frac{X^2}{\mu^2}}{H\left(M_{pl}^2+\xi\phi^2-15
\frac{X^2}{\mu^2}\right)+\gamma
X\dot{\phi}+\xi\phi\dot{\phi}}\right] \varTheta\,,
\end{eqnarray}
and
\begin{eqnarray}\label{55}
\partial^2{\cal{Q}}={\cal{X}}\dot{\varTheta}\,,
\end{eqnarray}
one can obtain the third order action up to the leading order (see
Appendix \textbf{B}). After obtaining the third order action
(\ref{56}), we are in the position to study the non-Gaussianity of
the primordial perturbations by evaluating the three-point
correlation functions. To this end, we use the interacting picture
and obtain the vacuum expectation value of the curvature
perturbation $\varTheta$ for the three-point operator in the
conformal time interval between the beginning of the inflation,
$\tau_i$, and the end of the inflation, $\tau_f$ as follows (see
for instance \cite{13,16,88})
\begin{eqnarray}\label{60}
\langle \varTheta(\textbf{k}_{1})\, \varTheta (\textbf{k}_{2})\,
\varTheta (\textbf{k}_{3})\rangle=
-i\int_{\tau_{i}}^{\tau_{f}}d\tau\,a\,\langle
0|[\varTheta({\tau_{f}},\textbf{k}_{1})\,\varTheta({\tau_{f}},\textbf{k}_{2})\,
\varTheta({\tau_{f}},\textbf{k}_{3})\,,\,
{\cal{H}}_{int}({\tau})]|0\rangle\,.
\end{eqnarray}
with ${\cal{H}}_{int}$ being the interacting Hamiltonian which is
equal to the Lagrangian of the cubic action. Since the coefficients
in the brackets of the Lagrangian (\ref{56}) vary slower than the
scale factor, we can approximate these coefficients to be constants
and solve the integral of equation (\ref{60}), which leads to the
following three-point correlation function of the curvature
perturbation in the Fourier space
\begin{eqnarray}\label{61}
\langle \varTheta(\textbf{k}_{1})\, \varTheta (\textbf{k}_{2})\,
\varTheta (\textbf{k}_{3})\rangle=
\big(2\pi\big)^{3}\delta^{3}\big(\textbf{k}_{1}+\textbf{k}_{2}+
\textbf{k}_{3}\big){\cal{A}}_s^2\,\,{\Xi}_{\varTheta}({\textbf{k}}_{1},{\textbf{k}}_{2},{\textbf{k}}_{3})\,,
\end{eqnarray}
with $A_s$ being the power spectrum of perturbation some time after
the Hubble radius crossing (given by equation (\ref{36})), and
${\Xi}$ being defined as
\begin{eqnarray}\label{62}
{\Xi}_{\varTheta}({\textbf{k}}_{1},{\textbf{k}}_{2},{\textbf{k}}_{3})=
\frac{(2\pi)^4}{\prod_{i=1}^{3}{\textbf{k}}_i^3}{\cal{G}}_\varTheta({\textbf{k}}_{1},{\textbf{k}}_{2},{\textbf{k}}_{3})\,.
\end{eqnarray}
Also, the parameter ${\cal{G}_{\varTheta}}$ is defined as
\begin{eqnarray}\label{63}
{\cal{G}}_{\varTheta}=
\frac{3}{4}\bigg(1-\frac{1}{c_{s}^{2}}\bigg)\,{\cal{Z}}_{1}
+\frac{1}{4}\bigg(1-\frac{1}{c_{s}^{2}}\bigg)\,{\cal{Z}}_{2}+
\frac{3}{2}\bigg(\frac{1}{c_{s}^{2}}-1-\frac{2\lambda}{\Sigma}+\Gamma\bigg)
\,{\cal{Z}}_{3}\,.
\end{eqnarray}
in which ${\cal{Z}}_1$, ${\cal{Z}}_2$ and ${\cal{Z}}_3$ are the
shape functions with the following relations
\begin{equation}\label{64}
{\cal{Z}}_{1}=\frac{2}{K}\sum_{i>j}k_{i}^{2}k_{j}^{2}-
\frac{1}{K^{2}}\sum_{i\neq j}k_{i}^{2}k_{j}^{3}\,,
\end{equation}
\begin{equation}\label{65}
{\cal{Z}}_{2}=\frac{1}{2}\sum_{i}k_{i}^{3}
+\frac{2}{K}\sum_{i>j}k_{i}^{2}k_{j}^{2}-\frac{1}{K^{2}} \sum_{i\neq
j}k_{i}^{2}k_{j}^{3}\,,
\end{equation}
\begin{equation}\label{66}
{\cal{Z}}_{3}= \frac{\big(k_{1}k_{2}k_{3}\big)^{2}}{K^{3}}\,,
\end{equation}
where
\begin{eqnarray}\label{67}
K=\sum_{i} k_{i}\,.
\end{eqnarray}
Equation (\ref{62}) shows that the three-point correlator depends
on the three momenta $k_1$, $k_2$ and $k_3$. We note that, in
order to satisfy the translation invariance, these momenta should
form a closed triangle with the constraint $k_1+k_2+k_3=0$
\cite{15,86,112,113,114}. Furthermore, considering the rotational
invariance makes the shape of the triangle important. Different
amount of momenta, gives different shapes of triangle and each
shape has a maximal signal in a special configuration. The
simplest one is the local shape \cite{115,116,117,118}, having a
peak in the squeezed limit with $k_3\rightarrow 0$ and $k_1\simeq
k_2$. There is another shape which corresponds to the equilateral
configuration \cite{15} with a signal at $k_1 = k_2 = k_3$. A
shape of non-Guassianity whose scalar product with the equilateral
shape vanishes, is called the orthogonal configuration \cite{119}.
A linear combination of the equilateral and orthogonal
configurations results in a shape corresponding to folded triangle
\cite{120} which is orthogonal to the equilateral templates and
has a maximal signal at the $k_1 = 2k_2 = 2k_3$ limit. We also
mention that, the orthogonal configuration has a signal with a
positive peak at the equilateral configuration and a negative peak
at the folded configuration. From the bispectrum
${\cal{G}}_{{\varTheta}}$ of the three-point correlation function
of curvature perturbations, the dimensionless parameter, $f_{NL}$,
characterizing the amplitude of non-Gaussianities, which is called
"nonlinearity parameter", is defined as
\begin{equation}\label{68}
f_{_{NL}}=\frac{10}{3}\frac{{\cal{G}}_{\varTheta}}{\sum_{i=1}^{3}k_{i}^{3}}\,.
\end{equation}
As has been noted previously, purely adiabatic Gaussian
perturbations result $f_{NL} = 0$, however, the presence of
non-Gaussian perturbations leads to deviation from $f_{NL} = 0$.
In what follows, we investigate the amplitude of non-Gaussianity
in the equilateral and orthogonal configurations. In this regard,
the bispectrum ${\cal{G}}_{\varTheta}$ in these configurations
should be obtained. To this end, we follow \cite{121,122,123} and
by considering
${\cal{G}}_{\varTheta}=\sum_{i=1}^{3}{\cal{G}}_{\varTheta}^{i}$,
we introduce the quantity ${\cal{I}}$ by the following expression
\begin{equation}\label{69}
{\cal{I}}(\Xi_{\varTheta}^{(i)}, \Xi_{\varTheta}^{(j)})=
\frac{{\cal{U}}(\Xi_{\varTheta}^{(i)}, \Xi_{\varTheta}^{(j)})}
{\sqrt{{\cal{U}}(\Xi_{\varTheta}^{(i)}, \Xi_{\varTheta}^{(i)})
{\cal{U}}(\Xi_{\varTheta}^{(j)}, \Xi_{\varTheta}^{(j)})}}\,,
\end{equation}
where
\begin{eqnarray}\label{70}
{\cal{U}}(\Xi_{\varTheta}^{(i)}, \Xi_{\varTheta}^{(j)})=\int
dk_{1}dk_{2}dk_{3}\Xi_{\varTheta}^{(i)}(k_1,k_2,k_3)
\Xi_{\varTheta}^{(j)}(k_1,k_2,k_3) \frac{(k_1 k_2
k_3)^4}{(k_1+k_2+k_3)^3}\,.
\end{eqnarray}
This integration should be done in the following region
\begin{eqnarray}\label{71}
0\leq k_{1}<\infty\,\,\,\,,\,\,\, 0<\frac{k_2}{k_1}<1
\,\,\,\,,\,\,\, 1-\frac{k_2}{k_1}\leq\frac{k_3}{k_1}\leq1\,.
\end{eqnarray}
For $|{\cal{I}}(\Xi_{\varTheta}^{(i)},
\Xi_{\varTheta}^{(j)})|\simeq 1$ the correlation is large, whereas
for $|{\cal{I}}(\Xi_{\varTheta}^{(i)},
\Xi_{\varTheta}^{(j)})|\simeq 0$ the two shapes are almost
orthogonal with a small correlation. Following \cite{122} we
define a shape ${{\cal{Z}}}_{*}^{equil}$ as
\begin{equation}\label{72}
{{\cal{Z}}}_{*}^{equil}=-\frac{12}{13}\Big(3{{\cal{Z}}}_{1}-{{\cal{Z}}}_{2}\Big)\,.
\end{equation}
Furthermore, we introduce another shape which is exactly orthogonal
to ${{\cal{Z}}}_{*}^{equil}$, as
\begin{equation}\label{73}
{{\cal{Z}}}_{*}^{ortho}=\frac{12}{14-13\beta}\Big[\beta\big(3{{\cal{Z}}}_{1}
-{{\cal{Z}}}_{2}\big)+3{{\cal{Z}}}_{1}-{{\cal{Z}}}_{2}\Big]\,,
\end{equation}
where $\beta=1.1967996$. Finally, using these relations, the
leading-order bispectrum (\ref{63}) can be written in terms of the
equilateral and orthogonal basis, ${{\cal{Z}}}_{*}^{equil}$ and
${{\cal{Z}}}_{*}^{ortho}$, as the following expressions
\begin{equation}\label{74}
{\cal{G}}_{\varTheta}={\cal{C}}_{1}\,{{\cal{Z}}}_{*}^{equil} +
{\cal{C}}_{2} \,{{\cal{Z}}}_{*}^{ortho}\,,
\end{equation}
with ${\cal{C}}_{1}$ and ${\cal{C}}_{2}$ being coefficients which
determine the magnitudes of the three-point correlation function
arising from equilateral and orthogonal contributions, respectively,
and are expressed as
\begin{eqnarray}\label{75}
{\cal{C}}_{1}=\frac{13}{12}\Bigg[\frac{1}{24}\left(1-\frac{1}{c_{s}^{2}}
\right)\left(2+3\beta\right)+\frac{\lambda}{12\Sigma}\big(2-3\beta\big)+
\left(\frac{1}{3{\cal{E}}_s
H\left(M_{pl}^2+\xi\phi^2+\frac{X^2}{\mu^2}\right)}
\right)\times\hspace{1cm}\\\nonumber \left(\gamma
X\dot{\phi}\left(\frac{2-3\beta}{2}-\frac{1}{c_s^2}\right)-6
H\frac{X^2}{\mu^2}\left(2-3\beta-\frac{1}{c_s^{2}}\right)\right)
\Bigg]\,,
\end{eqnarray}
and
\begin{eqnarray}\label{76}
{\cal{C}}_{2}=\frac{14-13\beta}{12}\Bigg[\frac{1}{8}\bigg(1-\frac{1}{c_{s}^{2}}\bigg)
-\frac{\lambda}{4\Sigma}+\frac{12 H\frac{X^2}{\mu^2}-\gamma
X\dot{\phi}}{2{\cal{E}}_s
H\Big(M_{pl}^2+\xi\phi^2+\frac{X^2}{\mu^2}\Big)}\Bigg]\,.
\end{eqnarray}
Here, $\lambda$ and $\Sigma$ are defined by equations (\ref{57}) and
(\ref{58}), respectively. By using equations (\ref{69})-(\ref{76}),
and also by definition of the non-linearity parameter (\ref{68}), we
find the following expressions for the amplitude of non-Gaussianity
in the equilateral and orthogonal configurations respectively
\begin{eqnarray}\label{77}
f_{_{NL}}^{equil}=\left(\frac{130}{36\sum_{i=1}^{3}k_{i}^{3}}\right)\Bigg[\frac{1}{24}\left(1-\frac{1}{c_{s}^{2}}
\right)\left(2+3\beta\right)+
\frac{\lambda}{12\Sigma}\left(2-3\beta\right)+\hspace{2cm}
\\\nonumber\Bigg(\gamma X\dot{\phi}
\left(\frac{2-3\beta}{2}-\frac{1}{c_s^2}\right)-\left(6
H\frac{X^2}{\mu^2}\right)
\left(2-3\beta-\frac{1}{c_s^{2}}\right)\Bigg)
\left(\frac{1}{3{\cal{E}}_s
H\left(M_{pl}^2+\xi\phi^2+\frac{X^2}{\mu^2}\right)}\right) \Bigg]
{{\cal{Z}}}_{*}^{equil}\,,
\end{eqnarray}
and
\begin{eqnarray}\label{78}
f_{_{NL}}^{ortho}=\bigg(\frac{140-130\beta}{36\,\sum_{i=1}^{3}k_{i}^{3}}\bigg)
\Bigg[\frac{1}{8}\bigg(1-\frac{1}{c_{s}^{2}}\bigg)
-\frac{\lambda}{4\Sigma}+ \frac{12 H\frac{X^2}{\mu^2}-\gamma
X\dot{\phi}}{2{\cal{E}}_s
H\Big(M_{pl}^2+\xi\phi^2+\frac{X^2}{\mu^2}\Big)}\Bigg]
{{\cal{Z}}}_{*}^{ortho}\,.
\end{eqnarray}
We again emphasize that , the shape function in the equilateral
configuration has a peak at the equilateral limit, ($k_1=k_2=k_3$).
Moreover, the orthogonal shape has a signal with a positive peak at
the equilateral configuration. Therefore, the nonlinearity parameter
in both configurations can be rewritten as
\begin{eqnarray}\label{79}
f_{_{NL}}^{equil}=\frac{325}{18}\Bigg[\frac{1}{24}\bigg(1-\frac{1}{c_{s}^{2}}
\bigg)\big(2+3\beta\big)+
\frac{\lambda}{12\Sigma}\big(2-3\beta\big)+ \Bigg(\gamma X\dot{\phi}
\left(\frac{2-3\beta}{2}-\frac{1}{c_s^2}\right)-\\\nonumber\left(6
H\frac{X^2}{\mu^2}\right)
\left(2-3\beta-\frac{1}{c_s^{2}}\right)\Bigg)
\Bigg(\frac{1}{3{\cal{E}}_s
H\Big(M_{pl}^2+\xi\phi^2+\frac{X^2}{\mu^2}\Big)}\Bigg) \Bigg]\,,
\end{eqnarray}
and
\begin{eqnarray}\label{80}
f_{_{NL}}^{ortho}=\frac{10}{9}\left(\frac{65}{4}\beta+\frac{7}{6}\right)
\Bigg[\frac{1}{8}\left(1-\frac{1}{c_{s}^{2}}\right)
-\frac{\lambda}{4\Sigma}+\frac{12 H\frac{X^2}{\mu^2}-\gamma
X\dot{\phi}}{2{\cal{E}}_s
H\left(M_{pl}^2+\xi\phi^2+\frac{X^2}{\mu^2}\right)}\Bigg]\,.
\end{eqnarray}
As Burrage et al. \cite{27} have mentioned, the non-Gaussianity is not constrained
to obey $f_{NL}\propto\frac{1}{c_{s}^2}$ in Galileon inflation as a new class of higher derivative
inflationary models. We see that this is
actually the case in our setup too. Non-Gaussinaties are not just
in the form of $f_{NL}\propto\frac{1}{c_{s}^2}$. As equations
(\ref{79}) and (\ref{80}) show, in the presence of higher order
derivatives of the Galileon field, non-gaussianities have more
complicated structure than the simple
$f_{NL}\propto\frac{1}{c_{s}^2}$ relation.

Up to this point, we have obtained the main equations of the model
at hand. In the next section, we test our inflationary model in
confrontation with Planck 2015 TT, TE, EE + low P and Planck2015
TTT, EEE, TTE and EET joint data to see the consistency and
viability of this model. We also find some constraints on the
model's parameters space (especially the non-minimal coupling
parameter) in this treatment.

\section{Observational Constraints}
In previous sections the primordial fluctuations in both linear
and nonlinear orders have been obtained. However, the viability of
an inflationary model depends on its perturbation parameters
coincidence with observational data. Thus, in what follows we find
some observational constraints on the parameters space of the
Galileon inflation treated in this paper. In this regard, we
should firstly define the form of the functions of the scalar
field introduced in action (\ref{action}), $V(\phi)$,
${\cal{K}}(\phi)$ and $\gamma(\phi)$. We adopt these functions as
$V(\phi)\sim\phi^m$, ${\cal{K}}(\phi)\sim\phi^m$ and
$\gamma(\phi)\sim\phi^m$ ($m=1$ corresponds to a linear potential,
$m=2$ refers to a quadratic potential and $m=3$ and $m=4$ imply
cubic and quartic potentials, respectively). After choosing these
forms of the generic functions, we analyze our model numerically
and find some constraints on the parameters space of the model.
For this purpose, at first, by substituting these functions into
the integral of equation (\ref{19}) and solving this integral, we
obtain the value of the inflaton field at the horizon crossing of
the physical scales in terms of the minimal number of e-folds,
$N$. Next, we use the obtained value of the field in equations
(\ref{37}), (\ref{49}), (\ref{53}), (\ref{79}) and (\ref{80}) in
order to calculate the spectral index of both scalar and tensor
perturbations modes, tensor-to-scalar ratio and the amplitudes of
the equilateral and orthogonal configurations of the
non-Gaussianity in terms of the minimal number of e-folds. Now, we
can explore the cosmological parameters numerically to see the
observational viability of this setup in confrontation with
recently released observational data. Another aspect of our study
is the deviation from the standard consistency relation of the
single-field inflation, $r=-8n_T$. Since Galileon inflation is a
generalized theory, we detect deviations away from the standard
consistency relation.

\begin{figure*}
\flushleft\leftskip0em{
\includegraphics[width=.41\textwidth,origin=c,angle=0]{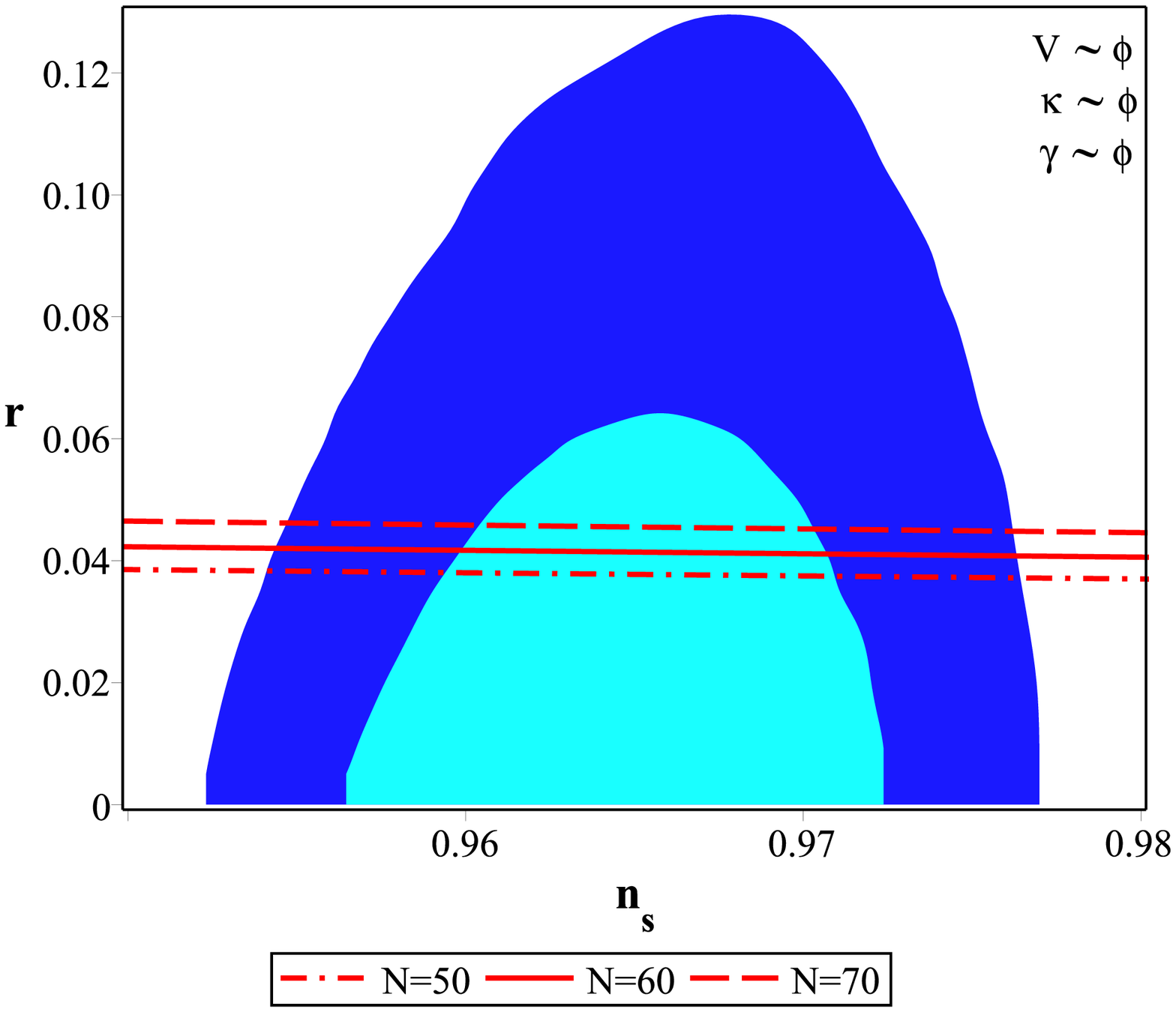}
\hspace{2cm}
\includegraphics[width=.42\textwidth,origin=c,angle=0]{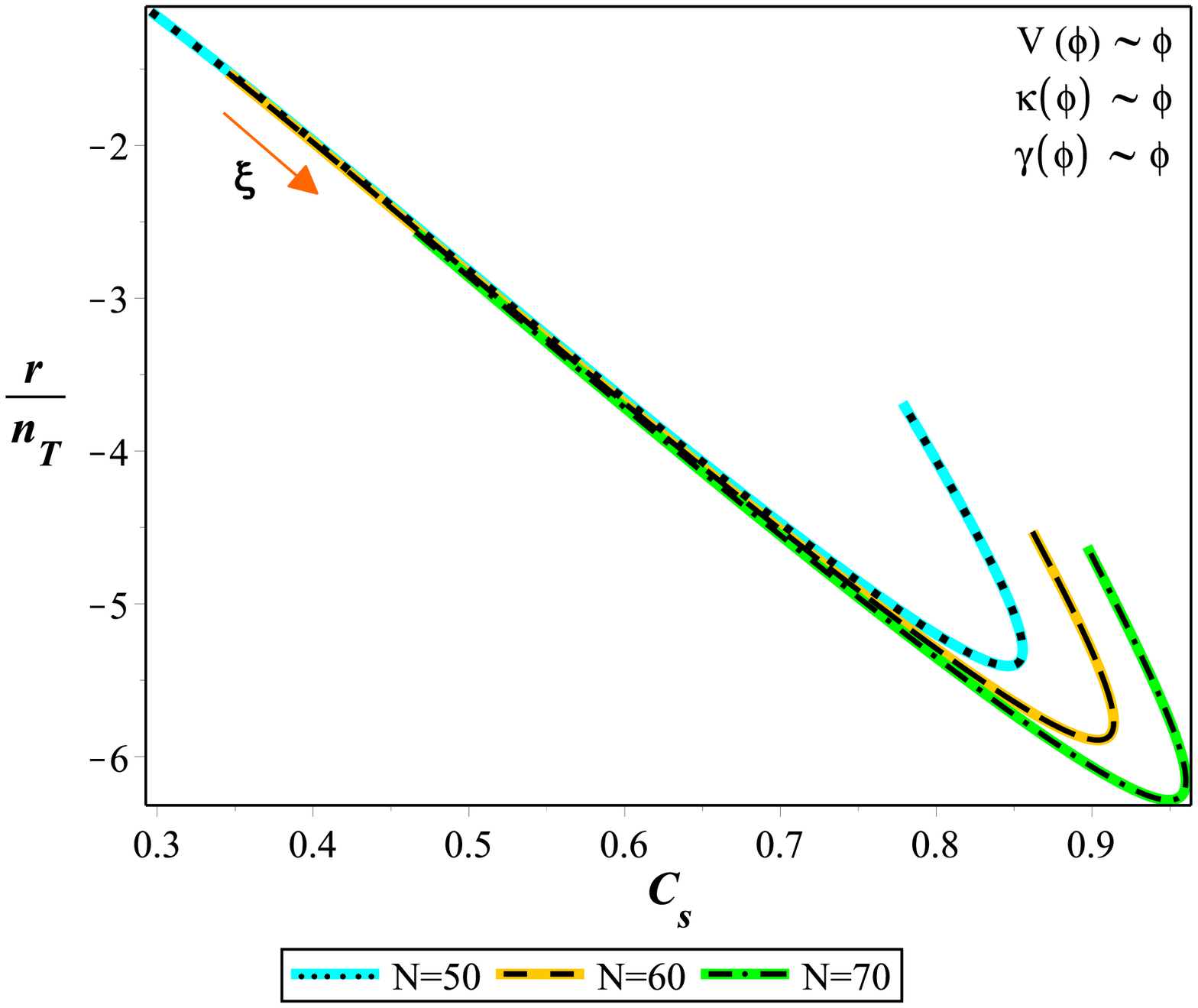}}
\caption{\label{fig:1} In the left panel the tensor-to-scalar ratio
versus the scalar spectral index in the background of the Planck2015
TT, TE, EE+lowP data for a Galileon inflationary model is plotted
with $m = 1$, which corresponds to potential of the form $V\sim\phi$
and also ${\cal{K}}\sim\phi$ and $\gamma\sim\phi$ for $N=50,\,60$
and $70$. Using the constraints obtained for $\xi$ in this case, the
ratio $\frac{r}{n_T}$ for the model is plotted as a function of the
sound speed, $c_s$, for each e-folds number in the right panel. This
figure shows deviation from the standard consistency relation
$r=-8n_T$.}
\end{figure*}

\begin{figure*}
\flushleft\leftskip0em{
\includegraphics[width=.41\textwidth,origin=c,angle=0]{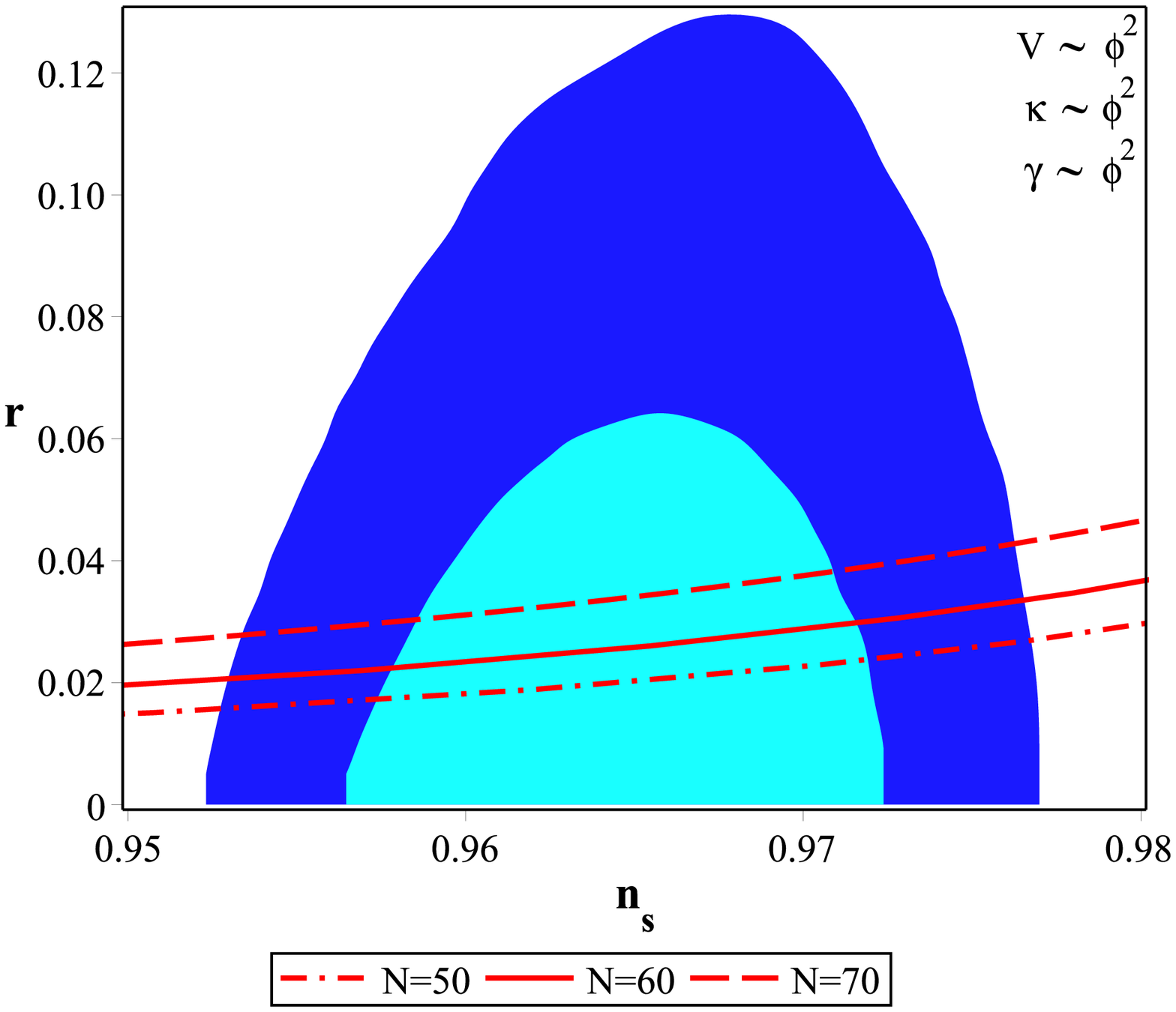}
\hspace{2cm}
\includegraphics[width=.42\textwidth,origin=c,angle=0]{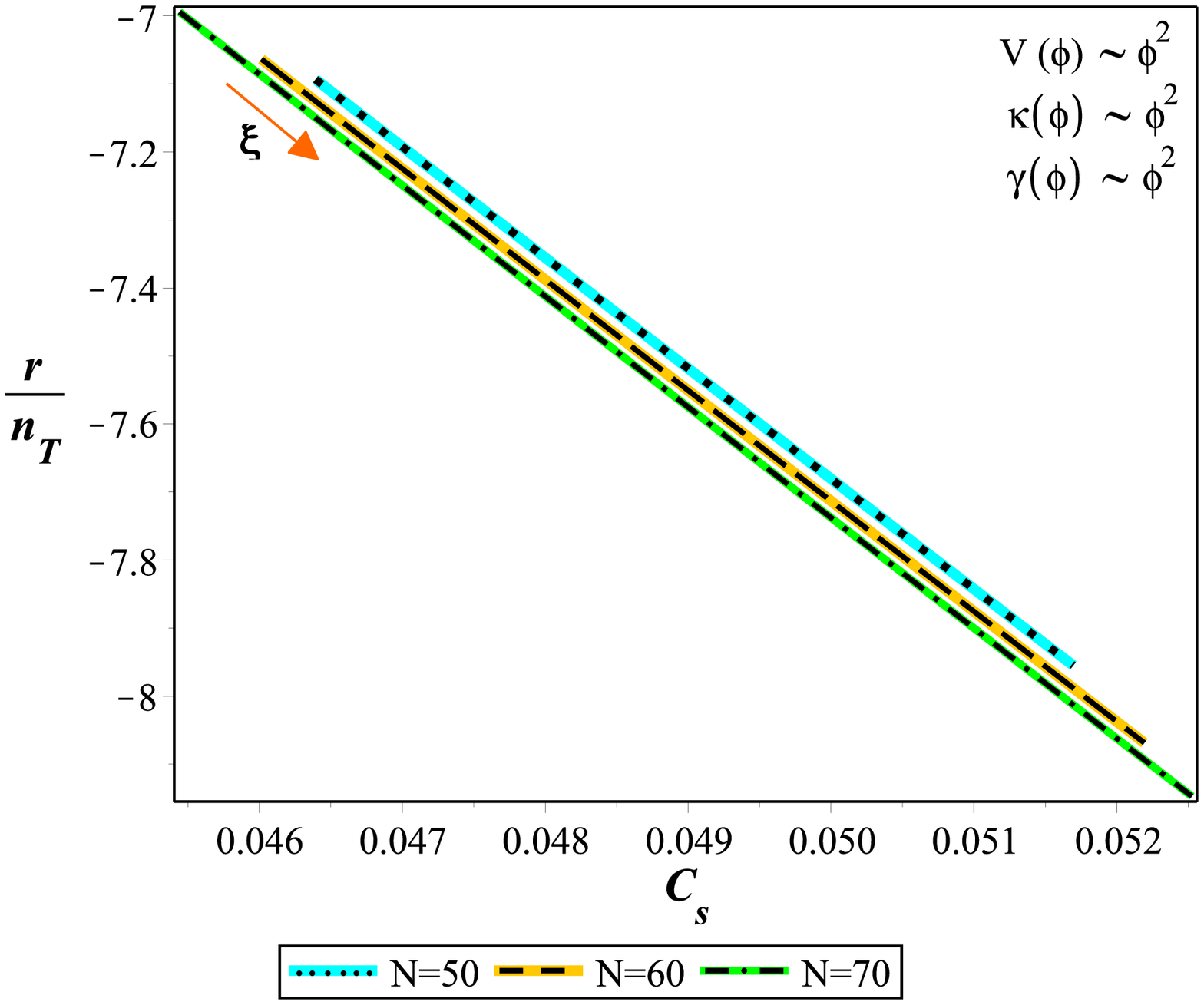}}
\caption{\label{fig:2} In the left panel the tensor-to-scalar ratio
versus the scalar spectral index in the background of the Planck2015
TT, TE, EE+lowP data for a Galileon inflationary model is plotted
with $m = 2$, which corresponds to potential of the form
$V\sim\phi^2$ and also ${\cal{K}}\sim\phi^2$ and $\gamma\sim\phi^2$
for $N=50,\,60$ and $70$. Using the constraints obtained for $\xi$
in this case, the ratio $\frac{r}{n_T}$ for the model is plotted as
a function of the sound speed, $c_s$, for each e-folds number in the
right panel. We find a \emph{slight} deviation from the standard
consistency relation $r=-8n_T$ which seems not to be enough
significant to be detected with confidence.}
\end{figure*}

\begin{figure*}
\flushleft\leftskip0em{
\includegraphics[width=.41\textwidth,origin=c,angle=0]{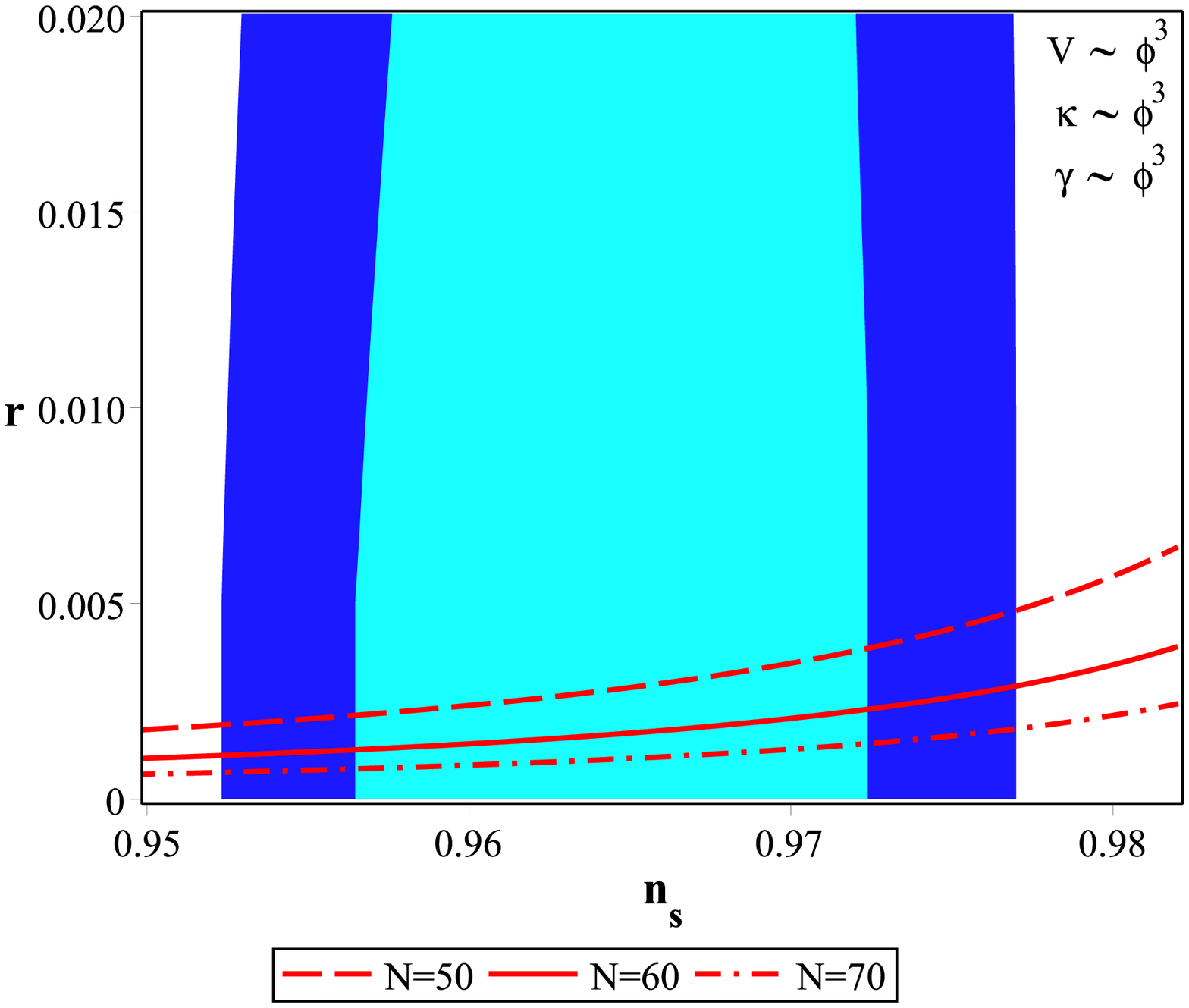}
\hspace{2cm}
\includegraphics[width=.42\textwidth,origin=c,angle=0]{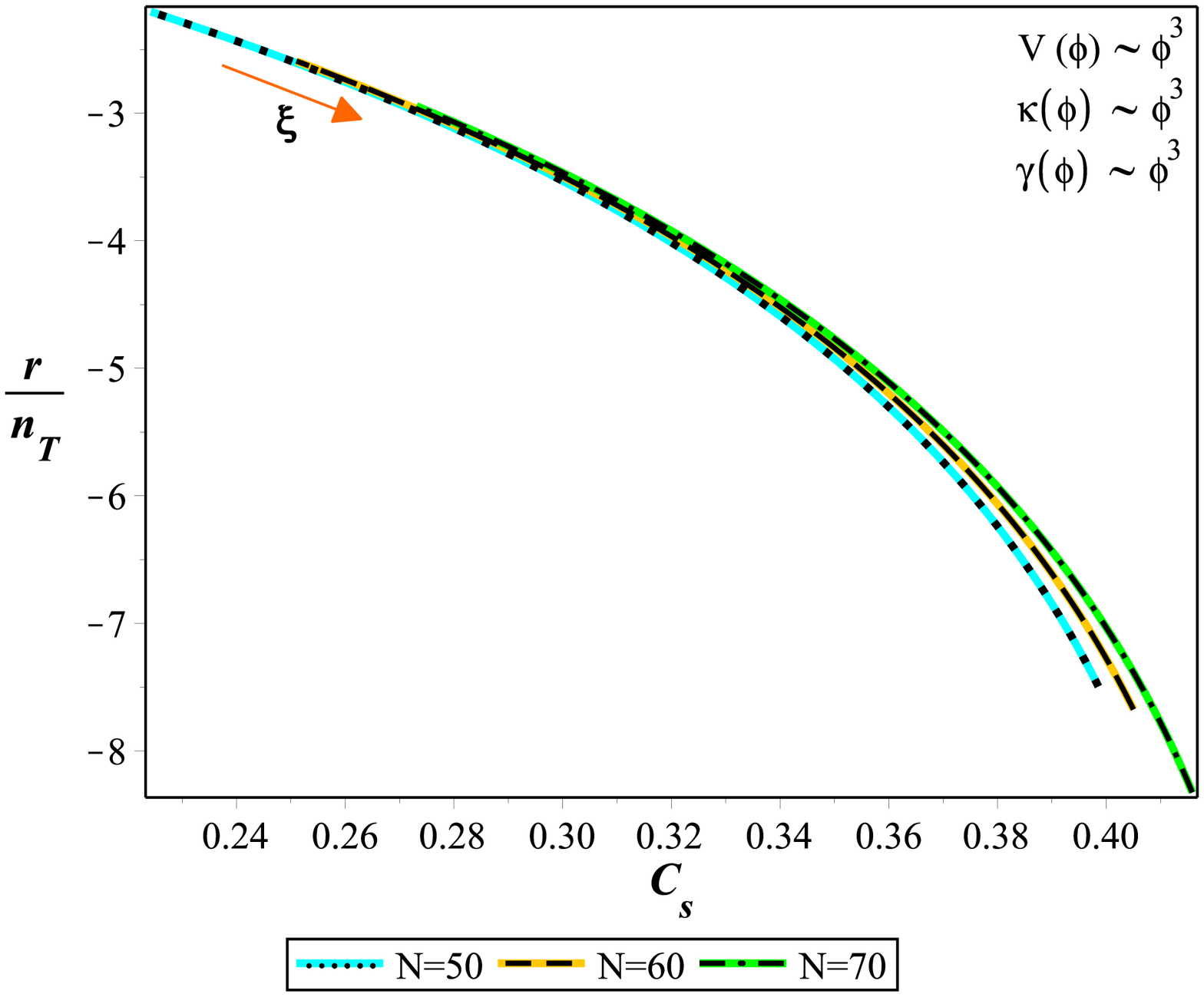}}
\caption{\label{fig:3} In the left panel the tensor-to-scalar ratio
versus the scalar spectral index in the background of the Planck2015
TT, TE, EE+lowP data for a Galileon inflationary model is plotted
with $m = 3$, which corresponds to potential of the form
$V\sim\phi^3$ and also ${\cal{K}}\sim\phi^3$ and $\gamma\sim\phi^3$
for $N=50,\,60$ and $70$. Using the constraints obtained for $\xi$
in this case, the ratio $\frac{r}{n_T}$ for the model is plotted as
a function of the sound speed, $c_s$, for each e-folds number in the
right panel. This figure shows deviation from the standard
consistency relation $r=-8n_T$.}
\end{figure*}

\begin{figure*}
\flushleft\leftskip0em{
\includegraphics[width=.41\textwidth,origin=c,angle=0]{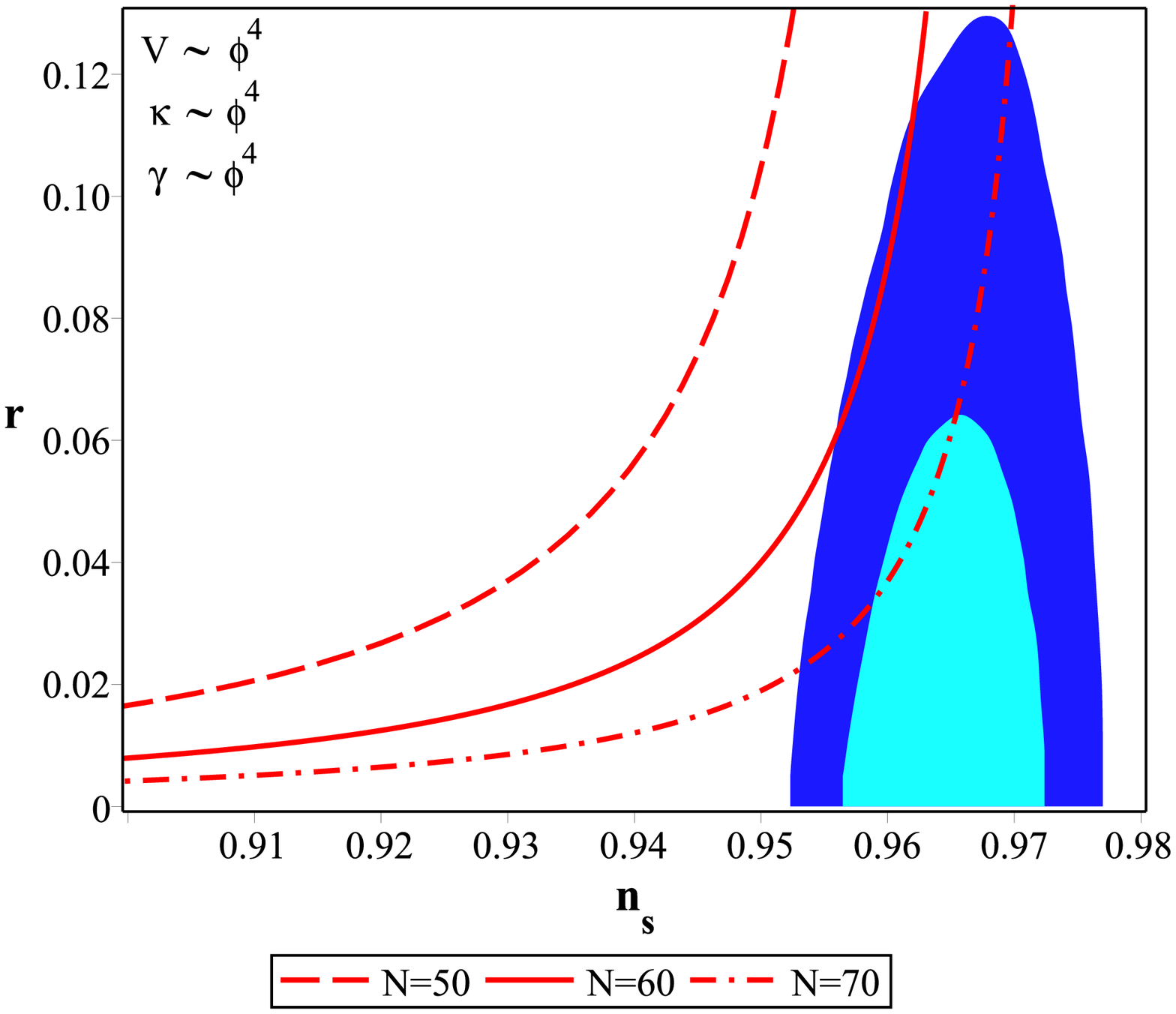}
\hspace{2cm}
\includegraphics[width=.42\textwidth,origin=c,angle=0]{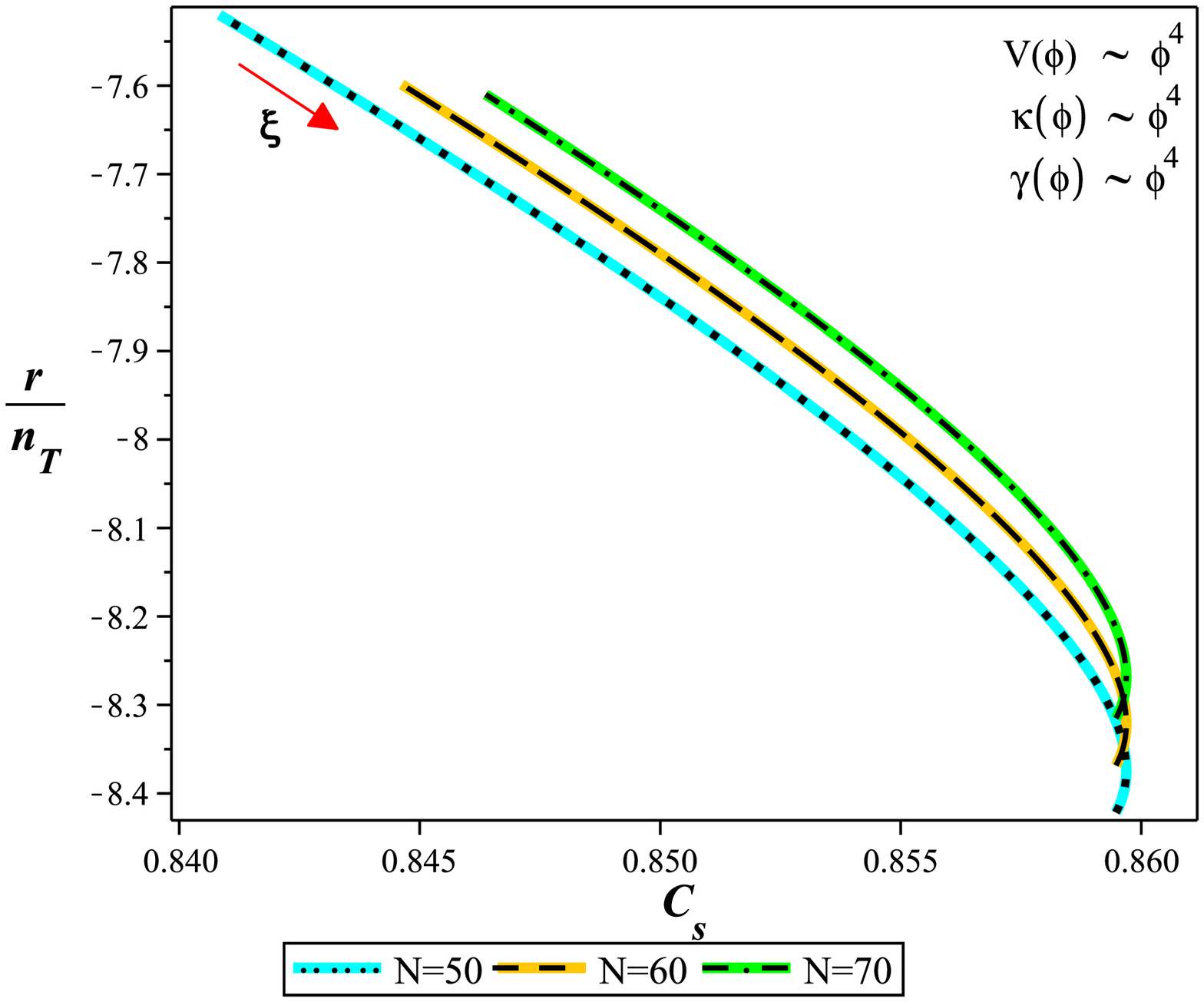}}
\caption{\label{fig:4} In the left panel the tensor-to-scalar ratio
versus the scalar spectral index in the background of the Planck2015
TT, TE, EE+lowP data for a Galileon inflationary model is plotted
with $m=4$, which corresponds to potential of the form $V\sim\phi^4$
and also ${\cal{K}}\sim\phi^4$ and $\gamma\sim\phi^4$ for
$N=50,\,60$ and $70$. Using the constraints obtained for $\xi$ in
this case, the ratio $\frac{r}{n_T}$ for the model is plotted as a
function of the sound speed, $c_s$, for each e-folds number in the
right panel. This figure shows \emph{slight} deviation from the
standard consistency relation $r=-8n_T$.}
\end{figure*}

\begin{figure*}
\flushleft\leftskip0em{
\includegraphics[width=.41\textwidth,origin=c,angle=0]{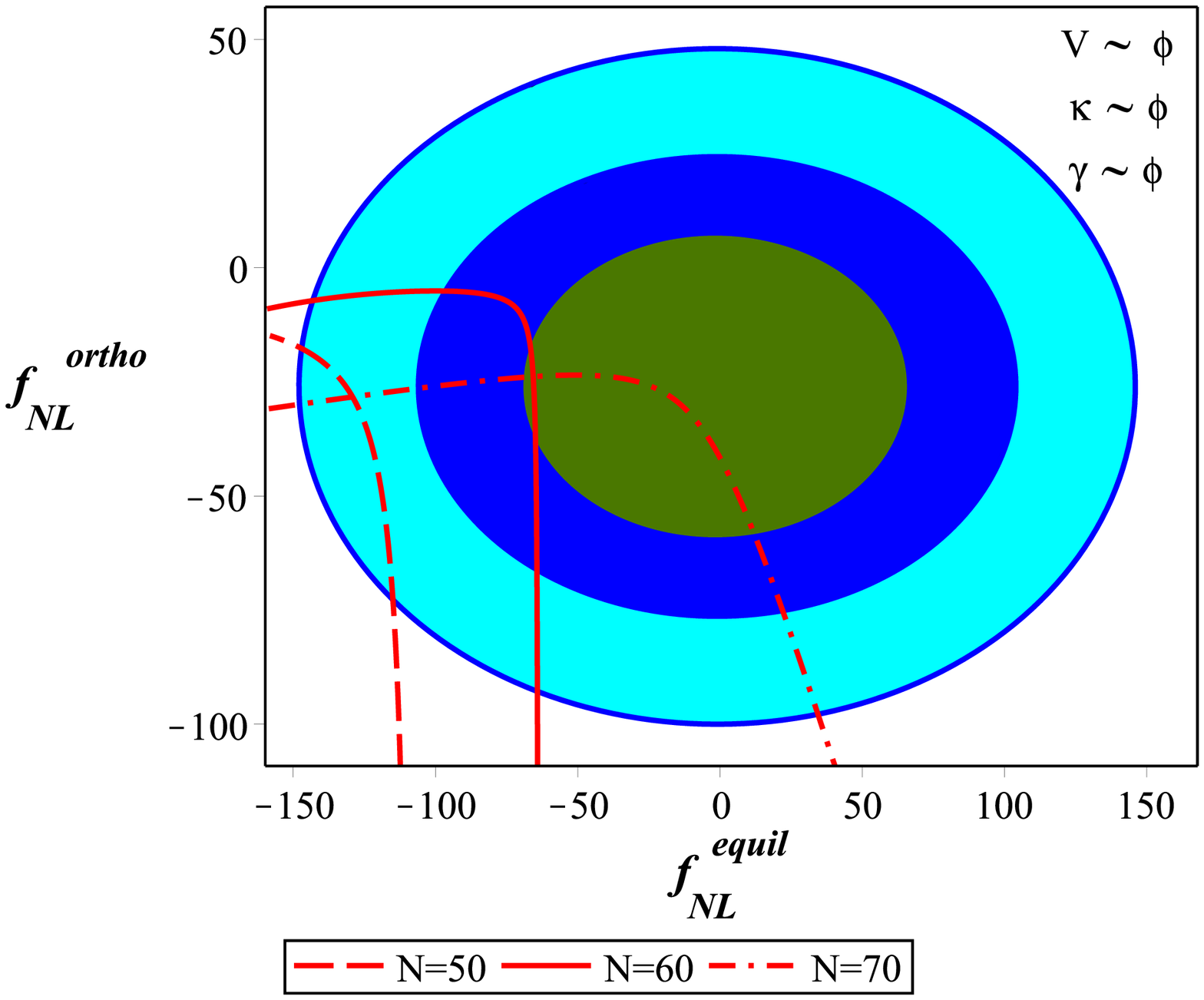}
\hspace{2cm}
\includegraphics[width=.42\textwidth,origin=c,angle=0]{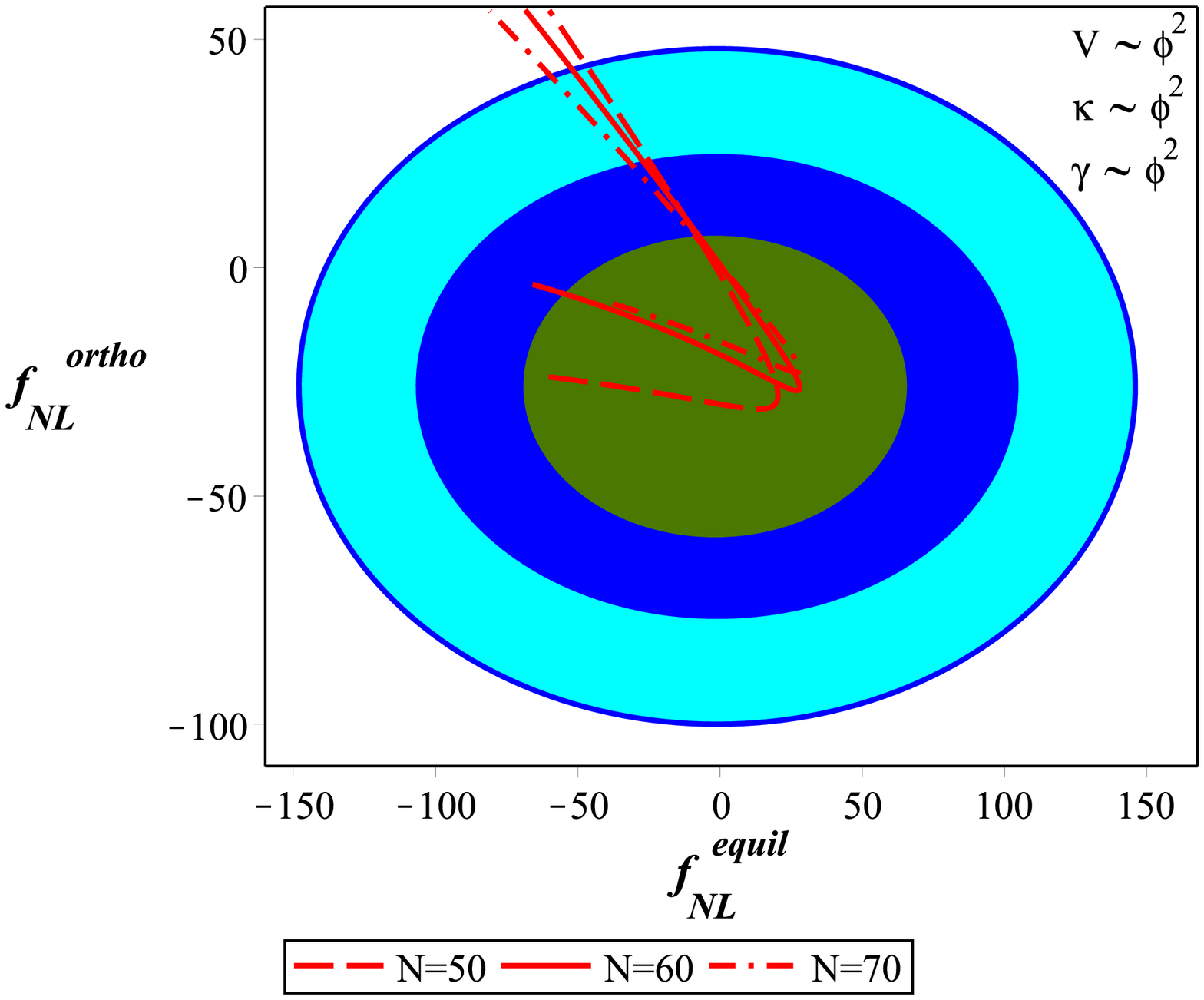}}
\caption{\label{fig:5} The amplitude of the orthogonal configuration
of the non-Gaussianity versus the amplitude of the equilateral
configuration for a Galileon inflationary model, in the background
of Planck2015 TTT, EEE, TTE and EET data. The left panel is plotted
for a linear potential with $m = 1$ while the right panel is plotted
for a quadratic potential with $m=2$. The figures are plotted in the
background of Planck2015 TTT, EEE, TTE and EET joint dataset and for
$N=50$, $60$ and $70$.}
\end{figure*}

\begin{figure*}
\flushleft\leftskip0em{
\includegraphics[width=.41\textwidth,origin=c,angle=0]{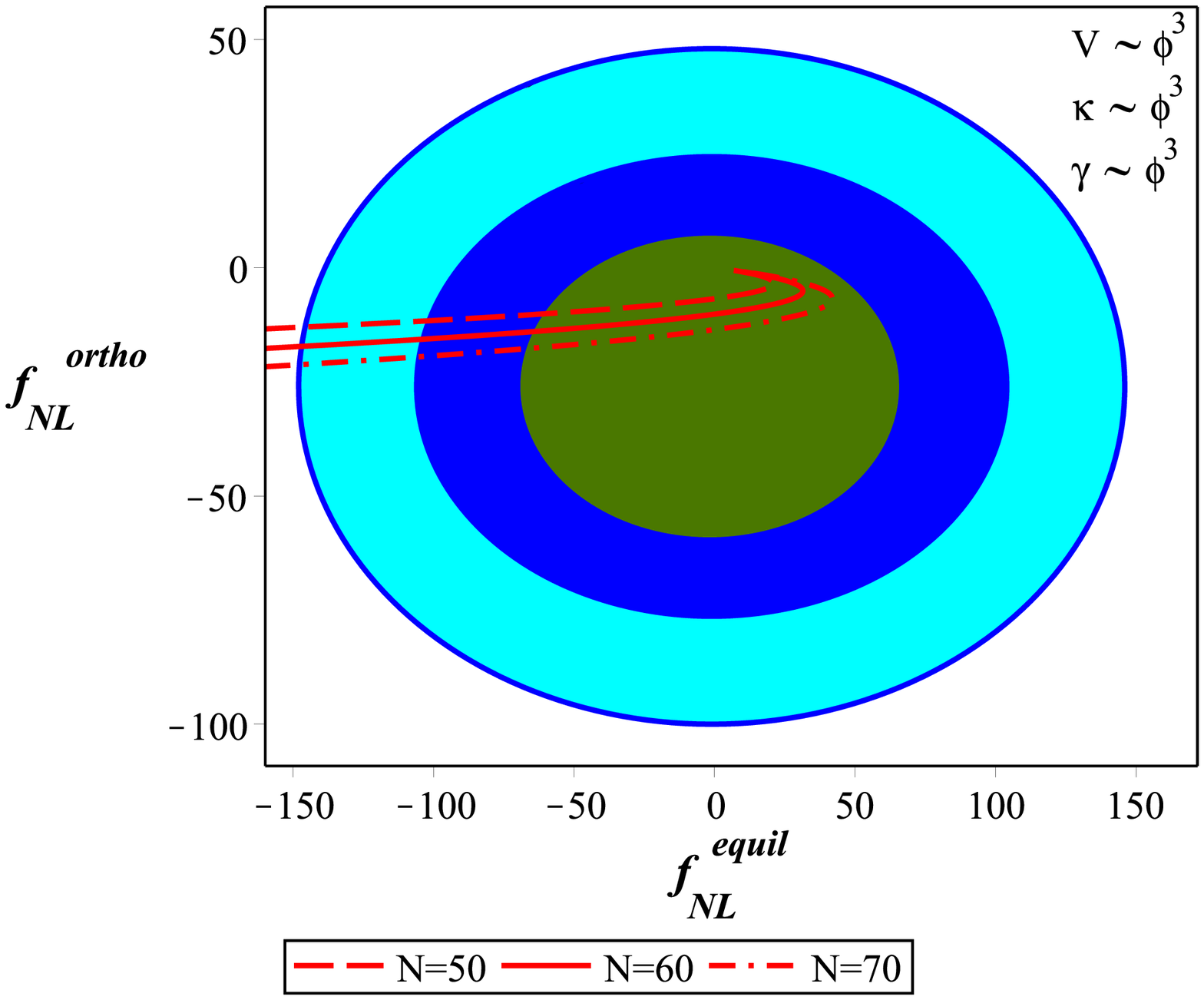}
\hspace{2cm}
\includegraphics[width=.42\textwidth,origin=c,angle=0]{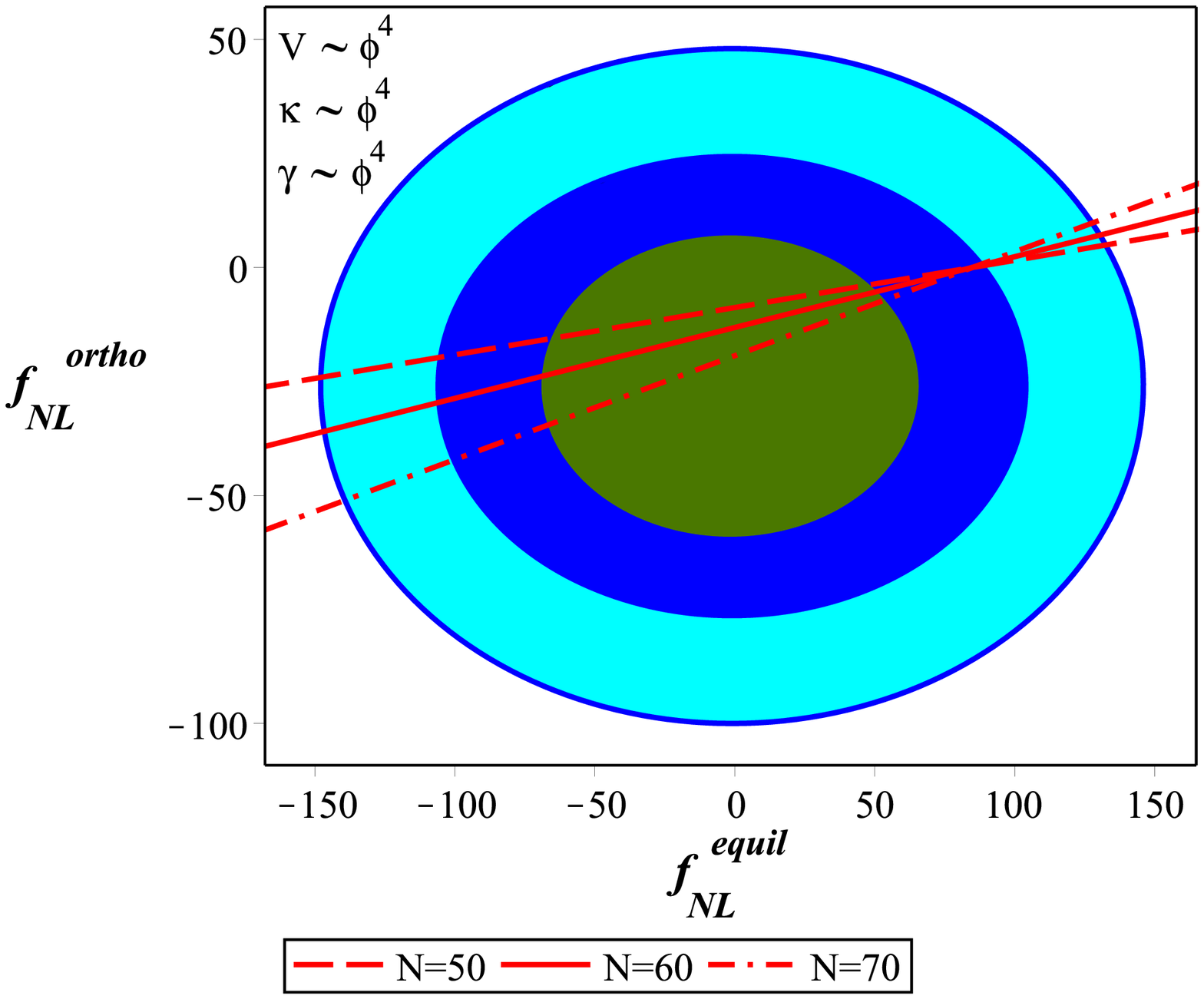}}
\caption{\label{fig:6} The amplitude of the orthogonal configuration
of the non-Gaussianity versus the amplitude of the equilateral
configuration for a Galileon inflationary model, in the background
of Planck2015 TTT, EEE, TTE and EET data. The right panel is plotted
for a cubic potential with $m = 3$ while the right panel is plotted
for a quartic potential with $m=4$. The figures are plotted in the
background of Planck2015 TTT, EEE, TTE and EET joint dataset and for
$N=50$, $60$ and $70$.}
\end{figure*}

Here we perform our analysis by setting $m=1,2,3,4$, (taking four
types of potentials), and $N=50$, $60$ and $70$ for each values of
$m$. The results are shown in figures
~\ref{fig:1},~\ref{fig:2},~\ref{fig:3},~\ref{fig:4},~\ref{fig:5}
and ~\ref{fig:6}. These figures confirm that a Galileon
inflationary model as described by the action (\ref{action}), with
nonzero $G_3$ and $G_4$ (which indicates the existence of a
non-minimal coupling between the scalar field $\phi$ and the Ricci
scalar $R$), in some ranges of the non-minimal coupling parameter,
$\xi$, is consistent with Planck2015 data. The ranges of the
acceptable values of the non-minimal coupling are shown in
tables~\ref{tab:1} and ~\ref{tab:2}.

Fig. ~\ref{fig:1} shows the behavior of the tensor-to-scalar ratio
versus the scalar spectral index (left panel) with $m=1$ (which
corresponds to $V(\phi)\sim\phi$, ${\cal{K}}\sim\phi$ and
$\gamma\sim\phi$) for $N=50$, $60$ and $70$ in the background of the
Planck2015 TT, TE, EE+lowP data. Our analysis shows that choosing a
linear form of the field for the functions $V$, ${\cal{K}}$ and
$\gamma$, a Galileon inflation is in a good agreement with recent
data if $0.0756845<\xi<0.0756980$ for $N=50$,
$0.0679086<\xi<0.067919$ for $N=60$ and $0.061955<\xi<0.0619636$ for
$N=70$. However, the right panel of this figure shows the ratio
$\frac{r}{n_T}$ of the model as a function of the sound speed $c_s$,
which shows deviation from the standard consistency relation,
$r=-8n_T$, for $m=1$.

In Fig. ~\ref{fig:2} the behavior of $r$ versus $n_s$ for $m=2$ is
plotted (left panel) which shows that adopting a quadratic form of
the functions  $V$, ${\cal{K}}$ and $\gamma$ is a suitable choice
for our Galileon inflation. Furthermore, as an important result, we
find a slight deviation from the standard consistency relation in
this case. Nevertheless, such a deviation seems not to be enough
significant to be detected with confidence.

In Fig. \ref{fig:3} we depicted our results for $m=3$ which is
consistent with observation for some ranges of $\xi$ that is given
in Table~\ref{tab:1}. The results of our analysis for the quartic
functions of the scalar field are drawn in Fig. \ref{fig:4}. This
figure shows that choosing $m=4$, the Galileon inflation is
consistent with recent data for $N=60$ and $N=70$. Deviation from
the standard slow-roll consistency relation is shown in the right
panel of each figures. One can see the ranges of the non-minimal
coupling parameter, $\xi$, in which the values of the inflationary
parameters $r$ and $n_s$ are compatible with the $95\%$ CL of the
Planck2015 TT, TE, EE+ low P joint data, in Table~\ref{tab:1}.

The numerical analysis on the non-Gaussian feature of the
perturbations in a Galileon inflation has been performed too. The
results are shown in Figs. \ref{fig:5} and ~\ref{fig:6}. As another
important result in our treatment, in some ranges of the NMC
parameter, it is possible to have large non-Gaussianity. For
instance, for $N=60$ and $m=2$, large non-Gaussianity can be
realized for region $\xi<0.00963$ in this setup. The ranges of the
non-minimal coupling parameter, $\xi$, in which the values of the
inflationary parameters $f^{ortho}_{NL}$ and $f^{equi}_{NL}$ are
compatible with the $95\%$ CL of the Planck2015 TTT, EEE, TTE and
EET joint data are given in Table~\ref{tab:2}.

\begin{table*}
\begin{center}
\caption{\label{tab:1}The ranges of the non-minimal coupling
parameter, $\xi$, in which the values of the inflationary parameters
$r$ and $n_s$ are compatible with the $95\%$ CL of the Planck2015
TT, TE, EE+low P joint data.}
\begin{tabular}{ccccc}
\\ \hline \hline$$& \,\,\,\,\,\,&$N=50$  &$N=60$&$N=70$\\ \hline\\
$m=1$& \,\,\,\,\,\,  &$0.0756845<\xi<0.0756980$&  \,\,\,\,\,\,\,$0.0679086<\xi<0.067919$& \,\,\,\,\,\,$0.061955<\xi<0.0619636$\\\\
$m=2$& \,\,\,\,\,\, &$0.007295<\xi<0.00769$&  \,\,\,\,\,\,\,$0.00675<\xi<0.007115$& \,\,\,\,\,\,$0.006325<\xi<0.00667$\\\\
$m=3$& \,\,\,\,\,\,& $0.009302<\xi<0.009535$&  \,\,\,\,\,\,\,$0.009162<\xi<0.009348$& \,\,\,\,\,\,$0.009054<\xi<0.009205$\\\\
$m=4$& \,\,\,\,\,\, & $not\,\,\,consistant$&  \,\,\,\,\,\,\,$0.02345<\xi<0.0249$& \,\,\,\,\,\,$0.0236<\xi<0.0285$\\ \hline\\\\
\end{tabular}
\end{center}
\end{table*}

\begin{table*}
\begin{center}
\caption{\label{tab:2}The ranges of the non-minimal coupling
parameter, $\xi$, in which the values of the inflationary parameters
$f^{ortho}_{NL}$ and $f^{equi}_{NL}$ are compatible with the $95\%$
CL of the Planck2015 TTT, EEE, TTE and EET joint data.}
\begin{tabular}{ccccc}
\\ \hline \hline$$& \,\,\,\,\,\,&$N=50$  &$N=60$&$N=70$\\ \hline\\
$m=1$& \,\,\,\,\,\,  &$0.06152<\xi<0.07619$&  \,\,\,\,\,\,\,$0.06142<\xi<0.08345$& \,\,\,\,\,\,$0.0603<\xi<0.08486$\\\\
$m=2$& \,\,\,\,\,\, &$\xi<0.0102$&  \,\,\,\,\,\,\,$\xi<0.00963$& \,\,\,\,\,\,$\xi<0.009384$\\\\
$m=3$& \,\,\,\,\,\,& $\xi<0.01563$&  \,\,\,\,\,\,\,$\xi<0.01553$& \,\,\,\,\,\,$\xi<0.01545$\\\\
$m=4$& \,\,\,\,\,\, & $0.01697<\xi<0.02872$&  \,\,\,\,\,\,\,$0.01723<\xi<0.0308$& \,\,\,\,\,\,$0.01785<\xi<0.03412$\\ \hline\\\\
\end{tabular}
\end{center}
\end{table*}

\section{Summary and Conclusions}

Successful inflationary models must obey an approximate shift
symmetry to derive enough e-folds of inflation.  Using this shift
symmetry we are allowed to add any scalar constructed from
gradients of the field to the inflationary Lagrangian. However, we
must be careful that adding arbitrary higher derivative operators
to the inflationary Lagrangian can lead to a loss of unitarity,
because of the appearance of ghost states \cite{67}. While one has
not to be worry about protection of the resulting Lagrangian from
large renormalizations, adding more input parameters than those
can be measured destroys the predictivity of the theory \cite{27}.

With these points in mind, in this paper we have studied a Galileon
inflation, which avoids the mentioned difficulties. We have
considered an inflationary Lagrangian containing non-canonical
derivative operators. The form of these operators are protected by
the covariant generalization of the Galileon shift symmetry. This
scenario contains a finite number of operators which lead to second
order field equations, implying the absence of ghosts.

We have firstly found main equations of the inflationary dynamics in
a Galileon inflation and then using the ADM formulation of the
metric, we have studied the linear perturbations in this setup. By
expanding the action up to the second order in perturbations in our
model, we have derived the two-point correlation functions and
obtained the amplitude of the scalar perturbation and its spectral
index. We have also derived the tensor perturbation of the model and
its spectral index by studying the tensor part of the perturbed
metric. The ratio between the amplitude of the tensor and scalar
perturbations (tensor-to-scalar ratio, $r$) has been obtained in
this setup. Furthermore, we have considered the consistency relation
in this Galileon model and found that  the consistency relation of
the standard inflation gets modified due to the presence of the
terms $G_3$ and $G_4$ in our action (\ref{action}). Finally, to
study the non-Gaussian feature of the primordial perturbations in
our model, we have studied the non-linear theory in details. In
order to investigate non-linear perturbation in the setup, one has
to expand the action up to the cubic order in perturbations and
obtain the three-point correlation functions. Hence, using the
interacting picture we have calculated the three-point correlation
functions and the nonlinearity parameter, $f_{NL}$, in this
generalized model. After introducing the shape functions as
${\cal{Z}}_{*}^{equil}$ and ${\cal{Z}}_{*}^{ortho}$, we have derived
the amplitude of non-Gaussianity in both equilateral and orthogonal
configurations. We have focused our attention on the equilateral
limit ($k_1=k_2=k_3$), in which, both the equilateral and orthogonal
configurations have peak.

After computing the main perturbation parameters, since the
viability of an inflationary model depends on its perturbation
parameters consistency with observational data, we have found some
observational constraints on the parameters space of the
inflationary model at hand. To this end, we have specified general
functions of the scalar field in the model as $V(\phi){\sim\phi^m}$,
${\cal{K}}\sim\phi^m$ and $\gamma\sim\phi^m$ with $m=1,2,3,4$. In
each case the behavior of the tensor-to-scalar ratio versus the
scalar spectral index for $N=50,\,60,$ and $70$ is depicted in
Figs.~\ref{fig:1}-~\ref{fig:4}. The results are summarized in
Table~\ref{tab:1}, which shows that for some ranges of the
non-minimal coupling parameter, $\xi$, the Galileon inflation is in
a good agreement with recent observation. Furthermore, an important
aspect of our study is the violation of the standard slow-roll
consistency relation of the single-field inflation, $r=-8n_T$.
Although we have observed from (\ref{53}) that the standard
consistency relation of the single-field inflation is in general
violated in this setup, however, we emphasize that, by choosing a
quadratic form for the scalar field functions, $V\sim\phi^2$,
${\cal{K}}\sim\phi^2$ and $\gamma\sim\phi^2$, we found a
\emph{slight} deviation from the standard consistency relation which
seems not to be enough significant to be detected confidently.

Moreover, the non-Gaussianity feature of the primordial
perturbations have been analyzed numerically by studying the
behavior of the orthogonal configuration versus the equilateral
configuration at the equilateral limit, $k_1=k_2=k_3$, and in the
background of the Planck2015 TTT, EEE, TTE and EET data. The
results are shown in figures (\ref{fig:5}) and (\ref{fig:6}). Also
the related constraints are presented in Table~\ref{tab:2}. As
another important result, we have shown that this Galileon
inflationary model allows to have large non-Gaussianity in some
ranges of the non-minimal coupling parameter that would be
observable by future improvements in experiments.\\

{\bf Appendix \textbf{A}: Expansion of the action up to the second order}\\

By replacing the perturbed metric (\ref{22}) in the action and
expanding the action up to the second order in perturbations,
the following expression will be obtained

\begin{eqnarray}\label{23}
\emph{S}_2=\int dt d^3x a^3
\Bigg[-3\left(M_{pl}^2+\xi\phi^2-3\frac{X^2}{\mu^2}\right)\dot{\varTheta}^2
-\frac{2}{a^2}\left(M_{pl}^2+\xi\phi^2-3\frac{X^2}{\mu^2}\right){\cal{B}}\partial^{2}\varTheta
+\hspace{1.5cm}\\\nonumber\frac{1}{a^2}\Bigg(\Big(2M_{pl}^2+2\xi\phi^2-
6\frac{X^2}{\mu^2}\Big) \dot{\varTheta}-
\left(2HM_{pl}^2+2H\Big(\xi\phi^2+\frac{X^2}{\mu^2}\Big) +2\gamma
X\dot{\phi}-32 H\frac{X^2}{\mu^2} +
2\xi\phi\dot{\phi}\right){\cal{B}}\Bigg)\partial^2\zeta+\\\nonumber
\Bigg(6H\Big(M_{pl}^2+\xi\phi^2+\frac{X^2}{\mu^2}\Big) +6\gamma
X\dot{\phi}-
96H\frac{X^2}{\mu^2}+6\xi\phi\dot{\phi}\Bigg){\cal{B}}\dot{\varTheta}+
\Bigg({\cal{K}}X-3H^2\left(M_{pl}^2+\xi\phi^2+\frac{X^2}{\mu^2}\right)-\\\nonumber
12\gamma HX\dot{\phi}+ 4\gamma_{,\phi}X^2+ 138
H^2\frac{X^2}{\mu^2}-6\xi H\phi\dot{\phi}\Bigg){\cal{B}}^2+
\frac{1}{a^2}\left(M_{pl}^2+\xi\phi^2+
\frac{X^2}{\mu^2}\right)\left(\partial\varTheta\right)^2 \Bigg]\,.
\end{eqnarray}

By varying this second order action with respect to the lapse
function and the shift vector, the equations of motion for
${\cal{B}}$ and $\zeta$ yield the constraints
\begin{eqnarray}\label{24}
{\cal{B}}=\left(\frac{M_{pl}^2+\xi\phi^2-3\frac{
X^2}{\mu^2}}{H\left(M_{pl}^2+\xi\phi^2-15 \frac{
X^2}{\mu^2}\right)+\gamma
X\dot{\phi}+\xi\phi\dot{\phi}}\right)\dot{\varTheta}\,,\hspace{1cm}
\end{eqnarray}
and
\begin{eqnarray}\label{25}
\frac{1}{a^2}\partial^2\zeta=3\dot{\varTheta}+\Bigg[-3H^2\left(M_{pl}^2+\xi\phi^2-45
\frac{ X^2}{\mu^2}\right)+ {\cal{K}}X-12\gamma
HX\dot{\phi}+4\gamma_{,\phi}X^2-6\xi
H\phi\dot{\phi}\Bigg]{\cal{B}}\times\hspace{0.7cm}\\\nonumber
\Bigg[{H\Big(M_{pl}^2+\xi\phi^2-15\frac{ X^2}{\mu^2}\Big)+ \gamma
X\dot{\phi}+\xi\phi\dot{\phi}}\Bigg]^{-1}-
\left[\frac{M_{pl}^2+\xi\phi^2-3\frac{X^2}{\mu^2}}{H\left(M_{pl}^2+
\xi\phi^2-15\frac{X^2}{\mu^2}\right) +\gamma X\dot{\phi}+
\xi\phi\dot{\phi}}\right]\frac{1}{a^2}\partial^2\varTheta\,.
\end{eqnarray}
Finally, substituting the equation of motion (\ref{24}) in the
second order action and taking some integrations by parts, we find
the quadratic action expressed in (\ref{26}).

\newpage
{\bf Appendix \textbf{B}: Expansion of the action up to the third order}\\

The third order action can be obtained as follows
\begin{eqnarray}\label{56}
S_3=\int dt d^3x
\Bigg\{a^3\Big(M_{pl}^2+\xi\phi^2+\frac{X^2}{\mu^2}\Big)
\Bigg[-\frac{3{\cal{E}}_s}{c_s^2} \Big(\frac{1}{c_s^2}-1\Big)
+\frac{{\cal{E}}_s}{c_s^4}\Big({\cal{E}}_s-\frac{\dot{{\cal{E}}_s}}
{H{\cal{E}}_s}\Big)+\hspace{1.5cm}\\\nonumber
\Bigg(\frac{{\cal{E}}_s}{c_s^4\Big(M_{pl}^2+\xi\phi^2+\frac{X^2}{\mu^2}\Big)}\Bigg)
\left(\frac{4\gamma
X\dot{\phi}}{H}-44\frac{X^2}{\mu^2}\right)\Bigg]
\varTheta\dot{\varTheta}^2+a\Big(M_{pl}^2+\xi\phi^2+\frac{X^2}{\mu^2}\Big)
\Bigg[{\cal{E}}_s\Big(\frac{1}{c_s^2}-1\Big)+\\\nonumber
\frac{{\cal{E}}_s}{c_s^2}\Bigg({\cal{E}}_s+\frac{\dot{{\cal{E}}_s}}{H{\cal{E}}_s}+
4\frac{X^2}{\mu^2\Big(M_{pl}^2+\xi\phi^2+\frac{X^2}{\mu^2}\Big)}
-\frac{2\dot{c_s}}{Hc_s}
\Bigg)\Bigg]\varTheta\Big(\partial\varTheta\Big)^2+a^3\left[\frac{{\cal{E}}_s}{M_{pl}Hc_s^2}\Big(\frac{1}{c_s^2}-1-
\frac{2\lambda}{\Sigma}+\Gamma\Big)\right]\times\\\nonumber
\Big(M_{pl}^2+\xi\phi^2+\frac{X^2}{\mu^2}\Big) \dot{\varTheta}^3-
\frac{2a^3{\cal{E}}_s}{c_s^2}\dot{\varTheta}\Big(\partial_i
\varTheta\Big)\Big(\partial_i{\cal{Q}}\Big)+
\left(\frac{a^3}{4\Big(M_{pl}^2+
\xi\phi^2+\frac{X^2}{\mu^2}\Big)}\right)\times\\\nonumber
\Bigg[{\cal{E}}_s+\frac{4\gamma X\dot{\phi}-8
H\frac{X^2}{\mu^2}}{H\Big(M_{pl}^2+\xi\phi^2+\frac{X^2}{\mu^2}\Big)}\Bigg]
\Big(\partial^2\varTheta\Big)\Big(\partial{\cal{Q}}\Big)-
\frac{2aM_{pl}}{H^2\Big(M_{pl}^2+\xi\phi^2+\frac{X^2}{\mu^2}\Big)}\left(4
H\frac{X^2}{\mu^2}-\gamma X \dot{\phi}\right)\times\\\nonumber
\Bigg[\partial^2\varTheta
(\partial_i\varTheta\Big)\Big(\partial_j{\cal{Q}}\Big)-\varTheta\partial_i\partial_j\Big(\partial_i\varTheta\Big)
\Big(\partial_j{\cal{Q}}\Big)\Bigg]-\frac{a}{H^3}\Bigg[12 H
\frac{X^2}{\mu^2}-2\gamma
X\dot{\phi}\Bigg]\dot{\varTheta}^2\Big(\partial^2\varTheta\Big)+\\\nonumber
\frac{2}{3a}\Bigg(\frac{\gamma X\dot{\phi}-6
H\frac{X^2}{\mu^2}}{H^3}\Bigg)
\Bigg[\partial^2\varTheta\Big(\partial\varTheta\Big)^2-
\varTheta\partial_i\partial_j\Big(\partial_i\varTheta\Big)\Big(\partial_j\varTheta\Big)\Bigg]\Bigg\}\,,
\end{eqnarray}

where the parameters $\lambda$, $\Sigma$ and $\Gamma$ are defined
by the following expressions
\begin{eqnarray}\label{57}
\lambda=\frac{1}{3M_{pl}^4}\Big(M_{pl}^2+\xi\phi^2+\frac{X^2}{\mu^2}\Big)^2\Bigg[4\gamma_{,\phi}X^2-3H\gamma
X\dot{\phi}+54 H^2\frac{X^2}{\mu^2}\Bigg]\,,
\end{eqnarray}

\begin{eqnarray}\label{58}
\Sigma=\frac{M_{pl}^2+\xi\phi^2-3\frac{X^2}{\mu^2}}{M_{pl}^4}\Bigg[3\left(HM_{pl}^2+
H\Big(\xi\phi^2+\frac{X^2}{\mu^2}\Big)+ \gamma X \dot{\phi}-16
H\frac{X^2}{\mu^2}
+\xi\phi\dot{\phi}\right)^2+\hspace{1.5cm}\\\nonumber
\Bigg(-3M_{pl}^2H^2-3H^2\Big(\xi\phi^2+\frac{X^2}{\mu^2}\Big)+
{\cal{K}}X- 12\gamma X H\dot{\phi}+ 4\gamma_{,\phi}X^2+138
H^2\frac{X^2}{\mu^2}-6\xi
H\phi\dot{\phi}\Bigg)\times\\\nonumber\left(M_{pl}^2+
\xi\phi^2-3\frac{X^2}{\mu^2}\right)\Bigg]\,,
\end{eqnarray}
and
\begin{eqnarray}\label{59}
\Gamma=\frac{1}{M_{pl}^2+\xi\phi^2+\frac{X^2}{\mu^2}}\Bigg\{\frac{\gamma
X\dot{\phi}}{H}\Big(3-\frac{1}{c_s^2}\Big)-12\frac{X^2}{\mu^2}
\Big(4-\frac{1}{c_s^2}\Big)+
\frac{\xi\phi\dot{\phi}}{H}\Big(1-\frac{1}{c_s^2}\Big)-\Bigg[\Big(\frac{\gamma
X \dot{\phi}}{H}\Big)^2+\hspace{1cm}\\\nonumber \frac{\gamma
X\dot{\phi}}{H}\left(\frac{\xi\phi\dot{\phi}}{H}-30\frac{X^2}{\mu^2}\right)
+136\frac{X^4}{H\mu^4}-\Big(\frac{X^2}{H\mu^2}\Big)
\left(\xi\phi\dot{\phi}-30 H \frac{X^2}{\mu^2}\right)\Bigg]
\left(\frac{6c_s^2}{{\cal{E}}_s\left(M_{pl}^2+\xi\phi^2+
\frac{X^2}{\mu^2}\right)}\right)\Bigg\}\,.
\end{eqnarray}

{\bf Acknowledgement} We thank Tsutomu Kobayashi for very constructive comments on original version of this work. We also appreciate an anonymous referee for
insightful comments which considerably improved the quality of the
paper.

%%%%%%%%%%%%%%%%%%%%%%%%%%%%%%%%%%%%%%%%%%%%%%%%%%%%%%%%%%%%%%%%%%%%%%%%%%%%%%%%%%%%%%%%%%%%%%%%%%%%%%%%%%%%%%
%%%%%%%%%%%%%%%%%%%%%%%%%%%%%%%%%%%%%%%%%%%%%%%%%%%%%%%%%%%%%%%%%%%%%%%%%%%%%%%%%%%%%%%%%%%%%%%%%%%%%%%%%%%%%%

\end{document}